\documentclass{aa}  

\usepackage{graphicx}
\usepackage{txfonts}
\usepackage[pagebackref=true]{hyperref}
\usepackage[usenames,dvipsnames]{xcolor}
\definecolor{DarkGreen}{rgb}{0,0.6,0}
\definecolor{xlinkcolor}{cmyk}{1,1,0,0}
 \hypersetup{ 
  pdfstartview={FitH},
  pdfnewwindow=true,     
  colorlinks=true,    
  linkcolor=xlinkcolor,    
  citecolor=xlinkcolor,    
  filecolor=xlinkcolor,  
  urlcolor=xlinkcolor,     
  final=true,
 }
\usepackage{booktabs}
\usepackage{multirow}

\usepackage{CJK}
\usepackage[utf8]{inputenc}
\usepackage{fontawesome}
\usepackage{lineno} 

\begin{document}
   \title{Active region upflows in various coronal structures and their coupling to the lower atmosphere}

\begin{CJK*}{UTF8}{gbsn}
   \author{Y.~Zhu (朱英杰)
          \inst{1,2}
          \and
          L.~Harra
          \inst{2,1}
          \and
          K.~Barczynski
          \inst{1,2}
          \and 
          N.~Janitzek
          \inst{1,2}
          \thanks{Now at European Space Agency, European Space Astronomy Center, Camino Bajo del Castillo, s/n Urbanizaci\'on Villafranca del Castillo, Villanueva de la Ca\~nada, 28692 Madrid, Spain}
          \and 
          J.~Plowman
          \inst{3}
          \and 
          S.~Mzerguat
          \inst{4}
          \and
          F.~Auch\`ere
          \inst{4}
          \and
          W.~T.~Thompson
          \inst{5}
          \and
          S.~Parenti
          \inst{2}
          \and
          L.~P.~Chitta
          \inst{6}
          \and
          H.~Peter
          \inst{6,7}
          \and
          T.~Fredvik
          \inst{8}
          \and
          T.~Grundy
          \inst{9}
          \and 
          Y.~W.~Ni (倪仪伟)
          \inst{10}
          \and \\
          P.~F.~Chen (陈鹏飞)
          \inst{10}
          \and 
          G.~Valori
          \inst{6}}

    \institute{ETH-Z\"urich, Wolfgang-Pauli-Str. 27, 8093 Z\"urich, Switzerland\\
    \email{yingjie.zhu@pmodwrc.ch}
    \and
        Physikalisch-Meteorologische Observatorium Davos/World Radiation Center, Dorfstrasse 33, 7260 Davos Dorf, Switzerland
    \and
        Southwest Research Institute, Boulder, CO 80302, USA
    \and
        Universit\'e Paris-Saclay, CNRS, Institut d'Astrophysique Spatiale, 91405 Orsay, France
    \and
        Adnet Systems, Inc., NASA Goddard Space Flight Center, Code 671, Greenbelt, MD 20771, USA
    \and
        Max Planck Institute for Solar System Research, Justus-von-Liebig-Weg 3, 37077 G\"ottingen, Germany
    \and 
        Institut f\"ur Sonnenphysik (KIS), Georges-K\"ohler-Allee 401A, 79110 Freiburg, Germany
    \and 
        Institute of Theoretical Astrophysics, University of Oslo, Oslo, Norway
    \and 
        RAL Space, UKRI STFC Rutherford Appleton Laboratory, Harwell, Didcot OX11 0QX, UK
    \and
        Key Laboratory of Modern Astronomy and Astrophysics, School of Astronomy and Space Science, Nanjing University, Nanjing 210023, PR China
    }

   \date{}

  \abstract
   {Plasma upflows with a Doppler shift exceeding $-10$\,km\,s$^{-1}$ at active region (AR) boundaries are considered potential sources of the nascent slow solar wind.  These upflows are often located at the footpoints of large-scale fan-like loops and show temperature-dependent Doppler shifts with redshifts in the transition region and blueshifts in the lower corona.}
   {We investigate the driving mechanisms of a pair of coronal upflow regions on the western and eastern peripheries of an AR, which have different magnetic topologies and surroundings. It is aimed to explore how these upflows couple to the lower atmosphere.}
   {Using observations of the Fe\,\textsc{xii} 19.51\,nm line from Hinode, we identified two upflow regions at the western and eastern boundaries of a decaying AR. Context images for the two regions were obtained by the High Resolution Imager (HRI) telescope of the Extreme Ultraviolet Imager (EUI) on board the Solar Orbiter mission. Other instruments on Solar Orbiter and other observatories provide diagnostics to the lower atmosphere. Potential Field Source Surface (PFSS) extrapolations were used to examine the magnetic field configuration associated with the AR upflows.}
   {The eastern upflow region, located over the AR moss, displays small-scale dynamic fibril structures, whereas the western region hosts fan-like loops. We found blueshifted Ne\,\textsc{viii} emission at the eastern site, in contrast to redshifted Ne\,\textsc{viii} profiles in the west. Magnetic field extrapolations reveal a pseudostreamer topology connecting both these regions. Moreover, low transition-region lines show systematically reduced redshift below the eastern footpoint.}
   {The observations support the scenario in which both upflows are driven by pressure imbalances created by coronal reconnection, leading to a continuous upflow above approximately 0.6\,MK (i.e., Ne\,{\sc viii} line formation temperature). Meanwhile, mass flows in the lower transition region beneath the eastern upflow region appear to respond passively to the pressure-driven coronal upflows.}
    
               
    \titlerunning{Upflows with Different Morphologies}
    \authorrunning{Y. Zhu et al.}
    
   \maketitle
   
\section{Introduction}

Persistent coronal upflows with blueshifts greater than $-10$\,km\,s$^{-1}$ are frequently observed at the boundary or periphery of solar active regions (ARs) by ultraviolet (UV) and extreme ultraviolet (EUV) spectrographs \citep[e.g.,][]{Brynildsen1998,Thompson2000,Sakao2007}. Over the past two decades, AR upflows have been extensively studied using the EUV Imaging Spectrograph \citep[EIS;][]{Culhane2007} on board the Hinode \citep{Kosugi2007} spacecraft \citep[see reviews by][]{Harra2012, HinodeReviewTeam2019, Tian2021}. These upflows are of significant interest, as they may serve as one of the potential source regions of the nascent slow solar wind \citep[e.g., ][]{Harra2008, Culhane2014, Zangrilli2016} and provide evidence for a ubiquitous mass circulation in the lower corona \citep{Marsch2008} and lower atmosphere \citep{McIntosh2009}. 

These upflows are commonly observed as fainter emission features at the peripheries of an AR, often located at the base of large-scale fan-like loops, particularly when low-latitude coronal holes are observed near the AR (e.g., \citealp[AR 10978:][]{Harra2008}; \citealp[AR 10926:][]{Doschek2007, DelZanna2008}; \citealp[AR 10938:][]{Hara2008}; and \citealp[AR 10942:][]{Sakao2007, Baker2009}). They also appear in other structures, for example, dark regions at the periphery of bright AR cores \citep[e.g.,][]{Scott2013, Baker2023} or even within AR cores \citep[e.g.,][]{Peter2010}. These upflows are typically observed in coronal emission lines formed at temperatures $>1$\,MK, showing blueshifts ranging from $-$10 to $-$50\,km\,s$^{-1}$ \citep[e.g.,][]{Baker2017}. On the other hand, redshifts of tens of kilometers per second are often found in lines forming at transition region temperatures ($<0.8$\, MK), which are often interpreted as the draining or cooling of coronal plasma \citep[e.g.,][]{DelZanna2008, Warren2011,Young2012}. In addition, Doppler shifts show a positive correlation with nonthermal broadening in upflows \citep{Doschek2008, Doschek2012}, which might result from a large dispersion in upflow speeds \citep{Demoulin2013}. Upflows are found to appear with flux emergence \citep{Harra2010} and persist for days \citep{Harra2017}. Moreover, there is no evidence of a change in flow speed with the AR age, as the variation in line-of-sight (LOS) velocity can be well explained by a steady-flow model \citep{Demoulin2013, Baker2017}. 

Potential driving mechanisms of AR upflows broadly fall into several, potentially coexisting categories: (1) interchange reconnection between close loops in AR core and open field lines or large loops, and (2) small-scale heating events at loop footpoints, including the direct heating of chromospheric plasma or gentle chromospheric evaporation due to coronal heating; (3) flow along open magnetic funnels \citep{Marsch2008}; (4) spectral signatures caused by slow magnetoacoustic waves \citep[e.g.,][]{Verwichte2010}; (5) a combination of the above mechanisms \citep[e.g.,][]{Barczynski2021}.

The coronal reconnection scenarios are strongly supported by the frequent overlap between upflow regions and the quasi-separatrix layers (QSLs), where reconnection is more likely to occur due to the rapid change in magnetic connectivity \citep[e.g.,][]{Baker2009, Mandrini2015, Edwards2016}. During AR expansion, reconnection between the dense, closed loops in the AR core and low-pressure ambient field lines creates a pressure imbalance in the reconnected field lines, driving upflows \citep{DelZanna2011, Bradshaw2011}. Other observational evidence includes: (a) high first ionization potential (FIP) bias in the upflow regions, suggesting the upflow plasma originated from closed fields \citep{Brooks2012, Brooks2015}; (b) continuous AR expansion as a potential driver of reconnection \citep{Harra2008, Murray2010}; (c) persistent radio bursts above the upflow ARs consistent with the continuous reconnection \citep{DelZanna2011, Harra2021}; (d) frequent appearances of upflows in pairs \citep{Baker2017}; (e) signatures of interchange reconnection in the middle corona over such ARs \citep[][]{2023NatAs...7..133C,2023SoPh..298...78W}.

Meanwhile, high-resolution imaging reveals various small-scale dynamics potentially driving upflows, which further suggests that they are caused by heating at coronal loop footpoints \citep[chromospheric evaporation, e.g.,][]{DelZanna2008} or even within the chromosphere and transition region \citep[e.g.,][]{McIntosh2009, Nishizuka2011, McIntosh2012}. Such dynamics include propagating disturbances \citep[PDs; e.g.,][]{Berghmans1999,DeMoortel2000,Sakao2007}, waves \citep[e.g.,][]{Nakariakov2000, Wang2009}, jets \citep[e.g.,][]{He2010}, spicules, and dynamic fibrils \citep[e.g.,][]{McIntosh2009, Harra2023}. Observations suggest some properties of the coronal upflows might be associated with the dynamics: for example, (a) weak blue asymmetry with a second blueshifted component of $-50$ to $-100$\,km\,s$^{-1}$ at loop footpoints \citep[e.g.,][]{Bryans2010} might be related to PDs \citep{Tian2011b, Wang2013}; (b) quasi-periodic oscillations in intensity, Doppler shifts, and line broadening \citep[e.g.,][]{Tian2011a, Nishizuka2011} can be driven by high-speed flows or magnetoacoustic waves related to dynamics such as type II spicules \citep{Tian2012}. Hence, the upflow plasma could evaporate as a result of heating in the lower corona \citep[e.g.,][]{Klimchuk2014} or be directly heated from the chromosphere \citep[e.g.,][]{DePontieu2011}. The coupling between the coronal upflows and their lower atmosphere is further highlighted by observations of a decrease in redshifts of transition region lines \citep[e.g.,][]{Polito2020} or patches of blueshifts in the chromospheric and transition-region beneath upflows \citep[e.g.,][]{Barczynski2021, Huang2021}. 

Despite extensive studies, no single mechanism fully accounts for all observational aspects of AR upflows. The persistent reconnection in the solar corona might have little influence on the small-scale dynamics in the lower corona, whereas direct heating of the chromospheric and transition region plasma may not supply enough mass flux into the coronal upflows \citep{Tripathi2013, Patsourakos2014} or their counterparts in the chromosphere \citep{Vanninathan2015}. The various observed properties of upflows may imply a combination of various driving mechanisms. For example, \citet{Peter2010} suggested the minor blueshifted component may be caused by the heating of individual loop strands, possibly appearing as type II spicules. In contrast, Doppler shifts in large-scale coronal structures are caused by siphon flows, loop draining, or open magnetic funnels. However, distinguishing various driving mechanisms remains a challenge, given the spatial resolution limitations of the current EUV spectrographs. High-resolution imaging might provide useful context information, particularly when spectroscopic observations with comparable spatial resolution are not available. For example, \citet{Brooks2020} made FIP bias measurements of upflows, using data from Hi-C 2.1 \citep{Rachmeler2019}, and suggested that there are two driving mechanisms (coronal reconnection and dynamic activity) in the upflows, which are unresolved in spectroscopic data with lower spatial resolution.

In this study, we continue to investigate the contribution of various driving mechanisms to upflows. Inspired by the insights from the Hi-C 2.1 observations \citep{Brooks2020}, we used observations from the High Resolution Imager (HRI) telescope of the Extreme Ultraviolet Imager \citep[EUI;][]{Rochus2020} on board the Solar Orbiter \citep{Muller2020}, which has higher spatial resolution near perihelion to resolve fine structures in upflows \citep[e.g.,][]{Harra2023, Barczynski2023}. Combined with spectroscopic observations spanning the chromosphere, transition region, and corona, we aim to address the following science questions: (a) Are AR upflows embedded in different structures driven by distinct mechanisms? (b) How does the coronal upflow couple with the low atmosphere under these driving mechanisms? The observations used by this study are summarized in Section~\ref{sec:method}. We present the results in Section~\ref{sec:res}. The implications for the driving mechanisms of AR upflows are discussed in Section~\ref{sec:dis}. 

\section{Observation overview}~\label{sec:method}
\begin{figure}[htb!]
    \centering
    \includegraphics[width=\linewidth]{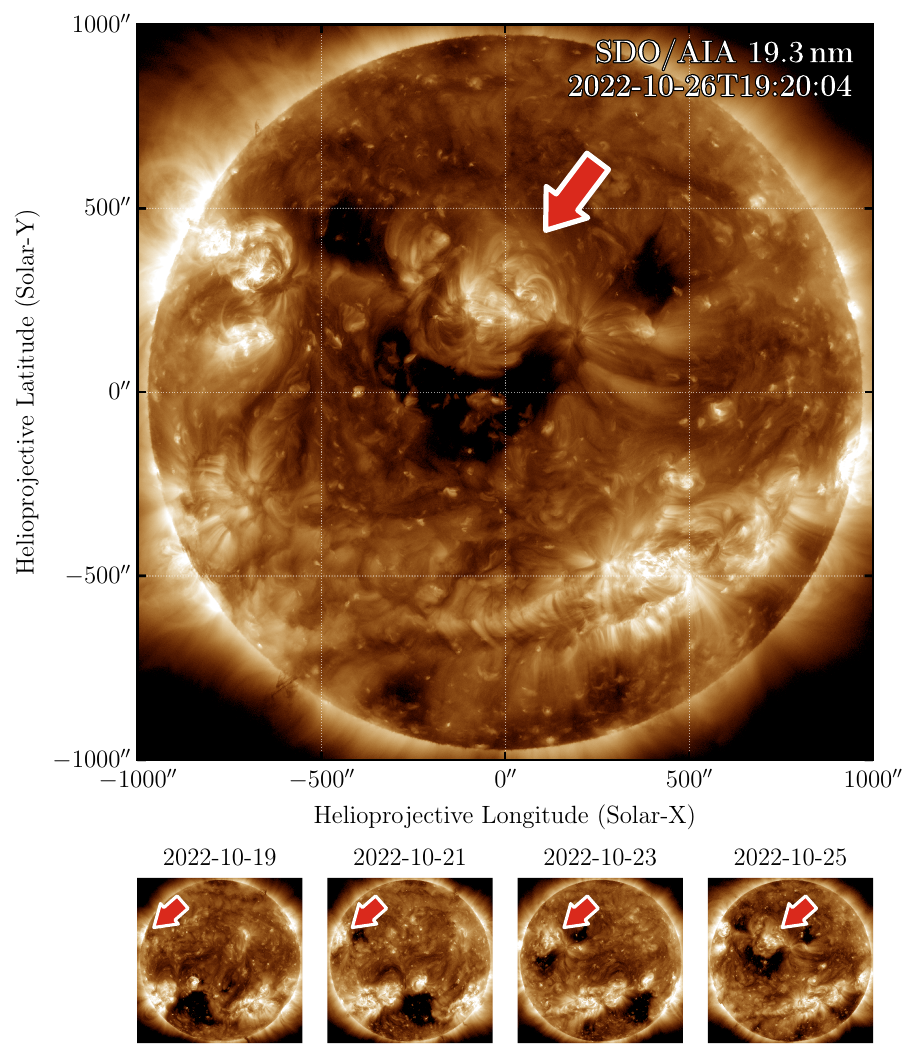}
    \caption{SDO/AIA 19.3\,nm images of the target AR (red arrow) between 2022 October 19 and October 26. Link to the \texttt{Jupyter} notebook creating this figure: \href{https://yjzhu-solar.github.io/EIS_DKIST_SolO/eis_eui_upflow_ipynb_html/halloween_smile.html}{\faBook}.}
    \label{fig:halloween_smile}
\end{figure}

The observations of upflow regions at the AR boundaries presented in this study were obtained by Solar Orbiter and other near-Earth observatories in the first coordinated observation campaign with the National Science Foundation's 4-m aperture Daniel K. Inouye Solar Telescope \citep[DKIST;][]{Rimmele2020} between 2022 October 17 and 2022 October 27. The Solar Orbiter observations were obtained as part of an AR\_Long\_Term SOOP \citep[Solar Orbiter Observing Plan;][]{Zouganelis2020}. We refer interested readers to \citet{Barczynski-submitted} for detailed information on this campaign. The observation target is a decaying AR (highlighted by an arrow in Figure~\ref{fig:halloween_smile}), which is the remnant of NOAA ARs 13110 and 13113 in the previous Carrington Rotation 2262. The target AR and three ambient low-latitude coronal holes form the ``Halloween Smiling Sun,'' with the AR as the ``nose'' of the ``smiling face.'' The ambient low-latitude coronal holes also make this decaying AR an interesting target to study the upflows, as previous studies suggested an enhancement of upflows and solar wind speeds when ARs sit beside the coronal holes \citep[e.g.,][]{Fazakerley2016}. 

\begin{figure*}[htb!]
    \centering
    \includegraphics[width=\linewidth]{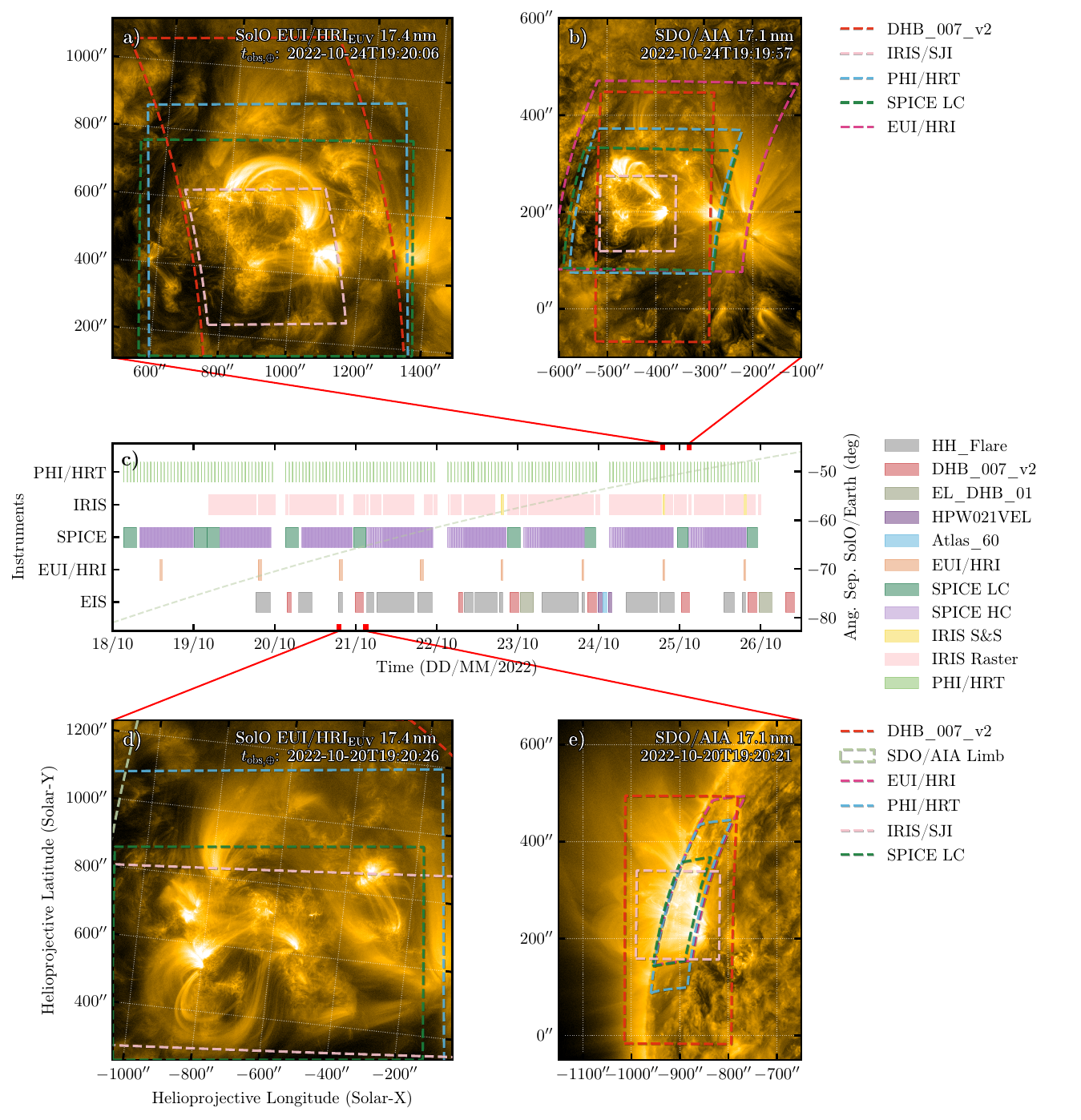}
    \caption{Timelines and FOVs of key observations analyzed in this study, when the target AR was seen from both Solar Orbiter and the Earth. (a) EUI/HRI context image at 2022 October 24 19:20:06 (light travel time corrected) with FOVs of Hinode/EIS, IRIS, PHI/HRT, and SPICE observations taken between October 24 19:00 and October 25 03:00 (red x-ticks in Panel (c)). (b) FOVs of observations shown in Panel (a), but reprojected to the perspective of SDO/AIA. (c) Timelines of IRIS, SPICE, EUI/HRI, and EIS observations, and the variation in the angular separation between Solar Orbiter and the Earth (dashed gray curve) from October 18 to October 26. Panels (d) and (e) are similar to Panels (a) and (b) but for observations Between October 20 19:00 to October 21 03:00. Acronyms in legends: HH\_Flare, DHB\_007\_v2, EL\_DHB\_01, HPW021VEL, and Atlas\_60 are various EIS studies names; LC and HC stand for low-cadence (long-exposure, large raster width) and high-cadence (short-exposure or narrow raster width) observations; S\&S denotes sit-and-stare observations. Link to the \texttt{Jupyter} notebook creating this figure: \href{https://yjzhu-solar.github.io/EIS_DKIST_SolO/eis_eui_upflow_ipynb_html/fov_summary.html}{\faBook}.}
    \label{fig:fov_summary}
\end{figure*}

Unfortunately, on-disk DKIST observations did not cover the upflow regions at the edge of the target AR due to a limited field of view (FOV). Therefore, we focused on spaceborne observations from Solar Orbiter and many other near-Earth observatories to study the plasma behavior in upflow regions and the underlying solar atmosphere. The key instruments on Solar Orbiter used in this study are the High Resolution Imager (HRI) and Full Disk Imager (FSI) telescopes of the Extreme Ultraviolet Imager \citep[EUI;][]{Rochus2020}, the Spectral Imaging of the Coronal Environment \citep[SPICE;][]{SPICEConsortium2020}, and the High Resolution Telescope \citep[HRT;][]{Gandorfer2018} of the Polarimetric and Helioseismic Imager \citep[PHI;][]{Solanki2020}. In addition, we also analyzed the observations made by the EUV Imaging Spectrometer \citep[EIS;][]{Culhane2007} on board the Hinode \citep{Kosugi2007} spacecraft, the Interface Region Imaging Spectrograph \citep{DePontieu2014,DePontieu2021}. Other full-disk synoptic observations made by the Atmospheric Imaging Assembly \citep[AIA;][]{Lemen2012} and the Helioseismic and Magnetic Imager \citep[HMI;][]{Scherrer2012} on board the Solar Dynamic Observatory \citep[SDO;][]{Pesnell2012}, and the data from the Chinese H$\alpha$ Solar Explorer \citep[CHASE;][]{Li2022} were also used in this study.

The week-long campaign provides an enormous dataset to study the evolution of the decaying AR and the associated upflows. We present the timeline of the various coordinated observations from the above instruments in Figure~\ref{fig:fov_summary}, especially the UV and EUV spectrographs and the high-resolution imaging instruments widely used in this study. The angular separation between the Sun-Solar Orbiter and the Sun-Earth lines is also presented in Figure~\ref{fig:fov_summary}c.  HRI\textsubscript{EUV} started to observe at 19:00 UT daily with a cadence of 5\,s. Between 2022 October 20 and 22, HRI\textsubscript{EUV} operated for one hour per day, while on the other days, it operated for half an hour. During the HRI\textsubscript{EUV} observation period, IRIS usually made high-cadence rasters or sit-and-stare observations, while dense 320-step rasters were made before or after HRI\textsubscript{EUV} observation. 

During most of the campaign, SPICE and EIS made high-cadence rasters over the AR core (e.g., SPICE study SCI\_AR-HEATING\_SC\_SL04\_9.7S\_FF and EIS study HH\_Flare+AR\_180x152H), which are designed for studies of AR evolution and flare eruption. Large-FOV rasters with longer exposure times (e.g., SPICE study SCI\_COMPO-TEST2\_SC\_SL04\_60.0S\_FF and EIS study DHB\_007\_v2), which provide better measurements of Doppler shifts, were run once or twice per day after the HRI\textsubscript{EUV} observation. These large rasters were also analyzed by Mzerguat et al. (in prep.) to study the AR elemental abundances. In addition, after daily coordination with HRI, SPICE often made a small raster scan with full-detector readouts (study name CAL\_SPECTRAL-RESPONSE\_FS\_SL04\_60.0S\_FD). As its narrow (120\arcsec$\times$660\arcsec) FOV missed the upflows, we used this dataset solely for the preliminary assessment of the SPICE point spread function (PSF, see more discussion in Section~\ref{subsec:spice_method} and Appendix~\ref{app:spice_psf}). 

Additionally, PHI/HRT took full-polarization images at a one-hour cadence almost uninterruptedly for the entire campaign, providing inverted quantities, including the photospheric magnetic field magnitude, inclination, and azimuth, continuum intensity, line-of-sight magnetic field and velocity, in addition to the measured full Stokes vectors of the Fe\,\textsc{i} 617.3\,nm line.

In this study, we primarily focused on observations obtained on 2022 October 20 and 24. Additional observations from other dates were also analyzed when they provided relevant information. The data calibration and coalignment are detailed in Appendices~\ref{app:data_calib} and \ref{app:image_reg}, respectively. For observations made by the Solar Orbiter instruments, all timestamps mentioned throughout the paper were adjusted to an Earth-based observer, with light travel time compensated (labeled as $t_\oplus$ or $t_{\mathrm{obs}, \oplus}$). 

Field of views of non-full-disk instruments on 2022 October 20 and October 24 are shown in the top and bottom panels of Figure~\ref{fig:fov_summary}, respectively, alongside contextual EUV images obtained by HRI\textsubscript{EUV} 17.4\,nm channel from the vantage point of Solar Orbiter, and AIA 17.1\,nm passband from the Earth's perspective. In HRI\textsubscript{EUV} and AIA images, the AR core exhibits a few closed loops and AR moss. A fan-like loop system is located at the western boundary of the AR. Unfortunately, most instruments, for example, HRI\textsubscript{EUV}, SPICE, and EIS, either missed or captured only a small portion of the fan loops during the campaign. The fan-like loop system was only well observed when the target AR was near the limb. 

\begin{figure*}[htb!]
    \centering
    \includegraphics[width=\linewidth]{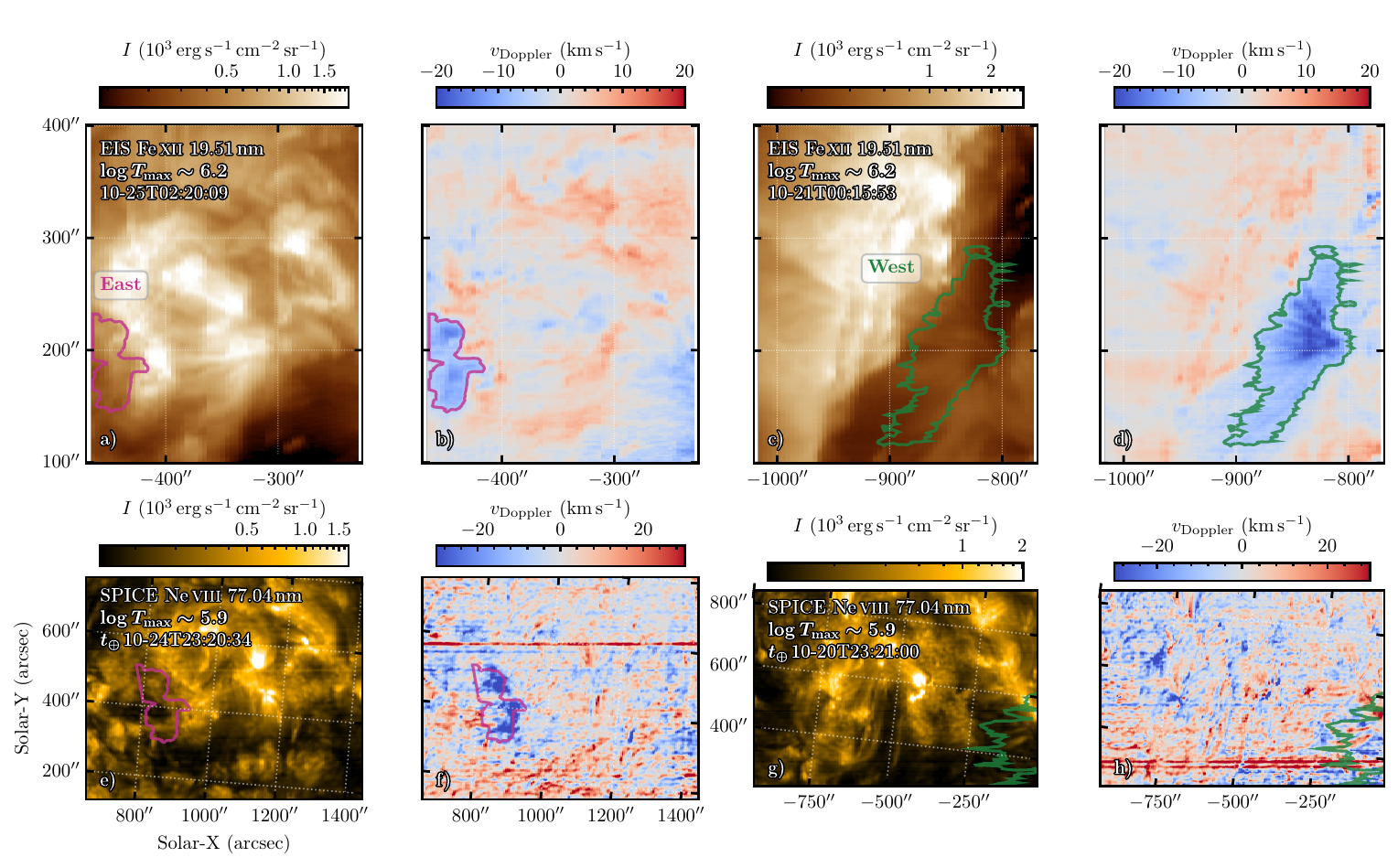}
    \caption{EIS and SPICE intensity and Doppler shift maps in Fe\,\textsc{xii} 19.51\,nm (EIS) and Ne\,\textsc{viii} 77.04\,nm lines of the eastern (purple) and western (green) upflow regions. Datasets obtained on different dates are shown due to the limited FOVs of EIS and SPICE. The eastern region was observed by EIS when the AR was near the disk center (e.g., 2022 October 24), while the western region was captured by EIS when the AR was close to the limb (e.g., October 20). Panels (a) and (b): Eastern upflow region in EIS; Panels (c) and (d): Western upflow region in EIS; Panels (e) and (f): Eastern upflow region in SPICE; Panels (g) and (h): Western upflow region in SPICE. Note that the spatial scales of SPICE and EIS in arcsec are different because Solar Orbiter was positioned at approximately 0.5\,AU from the Sun. Link to the \texttt{Jupyter} notebook creating this figure: \href{https://yjzhu-solar.github.io/EIS_DKIST_SolO/eis_eui_upflow_ipynb_html/eis_spice_doppler.html}{\faBook}.}
    \label{fig:eis_spice_doppler}
\end{figure*}

\section{Results}\label{sec:res}
\subsection{Temperature-dependent Doppler shifts}

Figure~\ref{fig:eis_spice_doppler} shows the intensity and Doppler velocity maps of the AR upflows on different dates. We first identified two major upflow regions on the eastern and western edges of the AR in the EIS Fe\,\textsc{xii} 19.51\,nm Dopplergrams. The eastern upflow region is close to the AR moss, exhibiting blurred emission in the Fe\,\textsc{xii} 19.51\,nm line. At lower temperatures (e.g., AIA 17.1\,nm), mossy structures are also found, but fainter than the moss in the AR core. Neither imaging nor spectroscopic observations show apparent coronal loops rooted in this upflow region. By contrast, the western upflow region is associated with large-scale fan-like loop systems, best seen in plasma emission around 1\,MK. 

By reprojecting the $-5$\,km\,s$^{-1}$ contours of Fe\,\textsc{xii} Doppler velocities to SPICE FOVs, we identified corresponding structures in the SPICE Ne\,\textsc{viii} Dopplergrams. In the eastern upflow region, patches of blueshifts greater than $-20$\,km\,s$^{-1}$ were found. The Ne\,\textsc{viii} blueshifts were concentrated in the western half of the Fe\,\textsc{xii} velocity contours, likely due to projection effects from the coronal Fe\,\textsc{xii} emission.  On the other hand, in the western upflow region, redshifts of tens of kilometers per second in Ne\,\textsc{viii} were found in fan-like loops and their footpoints, where the most prominent blueshifts in Fe\,\textsc{xii} are located. Additionally, cooler EIS lines, e.g., Fe\,\textsc{viii} and Si\,\textsc{vii}, also exhibit redshifts of 5--10\,km\,s$^{-1}$ in some fan-like loop strands. 

\begin{figure}[htb!]
    \centering
    \includegraphics[width=\linewidth]{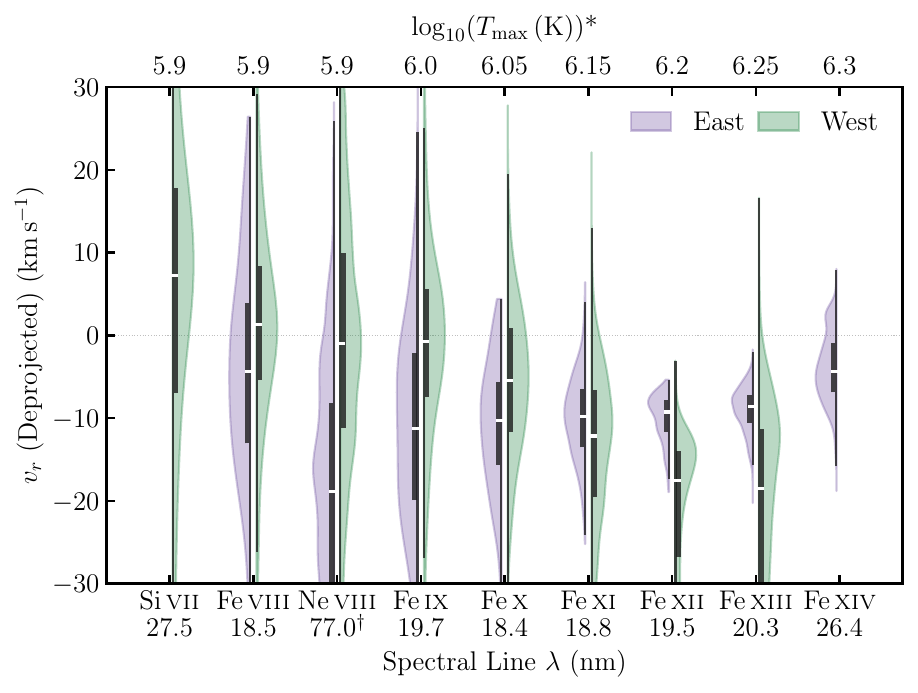}
    \caption{Violin plot comparing the kernel density distribution (KDE) of the deprojected Doppler shifts retrieved from the eastern (purple) and western (green) upflow regions outlined in Figure~\ref{fig:eis_spice_doppler}. Negative values represent blueshifts, while positive values stand for redshifts. The spectral lines are ordered by their formation temperatures. The thick black bars indicate the interquartile ranges, and the white dots represent the median of the distribution. *The maximum formation temperatures $T_{\mathrm{max}}$ of spectral lines calculated by an AR DEM derived using \citet{Vernazza1978} composite AR spectra. \textsuperscript{\textdagger}Ne\,\textsc{viii} Doppler shifts were measured by SPICE without additional corrections. Link to the \texttt{Jupyter} notebook creating this figure: \href{https://yjzhu-solar.github.io/EIS_DKIST_SolO/eis_eui_upflow_ipynb_html/eis_spice_doppler_vs_tmax.html}{\faBook}.}
    \label{fig:doppler_vs_tmax}
\end{figure}

To investigate the temperature dependence of Doppler shifts in the upflow region, we show the kernel density distribution (KDE) of Doppler shifts in spectral lines with various formation temperatures in Figure~\ref{fig:doppler_vs_tmax}. Because observations of the western and eastern upflow regions were made from different vantage points, velocities are deprojected, assuming that upflows are radial. Different temperature dependences of Doppler shifts were found in the two upflow regions. The western upflow region shows a typical pattern of Doppler shifts in fan loops -- redshifts in the upper transition region temperature (below 1\,MK) and a gradual transition into blueshifts in the corona above 1\,MK. The median Doppler velocity varies from 10\,km\,s$^{-1}$ in redshift to $-20$\,km\,s$^{-1}$ in blueshift. In contrast, blueshifts greater than 10\,km\,s$^{-1}$ were observed in the eastern upflow region from Fe\,\textsc{viii} and Ne\textsc{viii}) lines, implying a dominance of upward motion for plasma between 0.6 and 1\,MK. In the AR corona above 1.5\,MK, the median blueshifts are approximately $-20$\,km, s$^{-1}$ in the western fan loops, which are slightly greater than those of around $-10$\,km\,s$^{-1}$ observed in the eastern upflow regions.   

Due to the limited signal-to-noise ratio (S/N), it is challenging to make robust double-Gaussian fitting or the red-blue (RB) asymmetry analysis \citep{DePontieu2009a, Tian2011b} of line profiles in these EIS datasets. We attempted to perform the RB analysis on the brightest Fe\,\textsc{xii} 19.5119\,nm line using the modified technique \citep{Tian2011b} between Doppler shifts of 60 and 120\,km\,s$^{-1}$ away from the line core. In the eastern upflow region observed on 2022 October 24, a minor blue asymmetry of approximately $-$0.03 was found, much less compared to values of 0.10--0.20 \citep[e.g.,][]{Tian2011b}. This weak blue asymmetry could be affected by the low S/N and projection effects. Fe\,\textsc{xii} line profiles in other regions show a red asymmetry of 0.1, which could be caused by the weak Fe\,\textsc{xii} 19.5179\,nm blended in the red wing. The blended line is too weak in most pixels to fit with a Double Gaussian function. We also performed RB analysis on the western upflow region observed on October 21, but the results were too noisy to reach a solid conclusion.

\begin{figure*}[htb!]
    \centering
    \includegraphics[width=\linewidth]{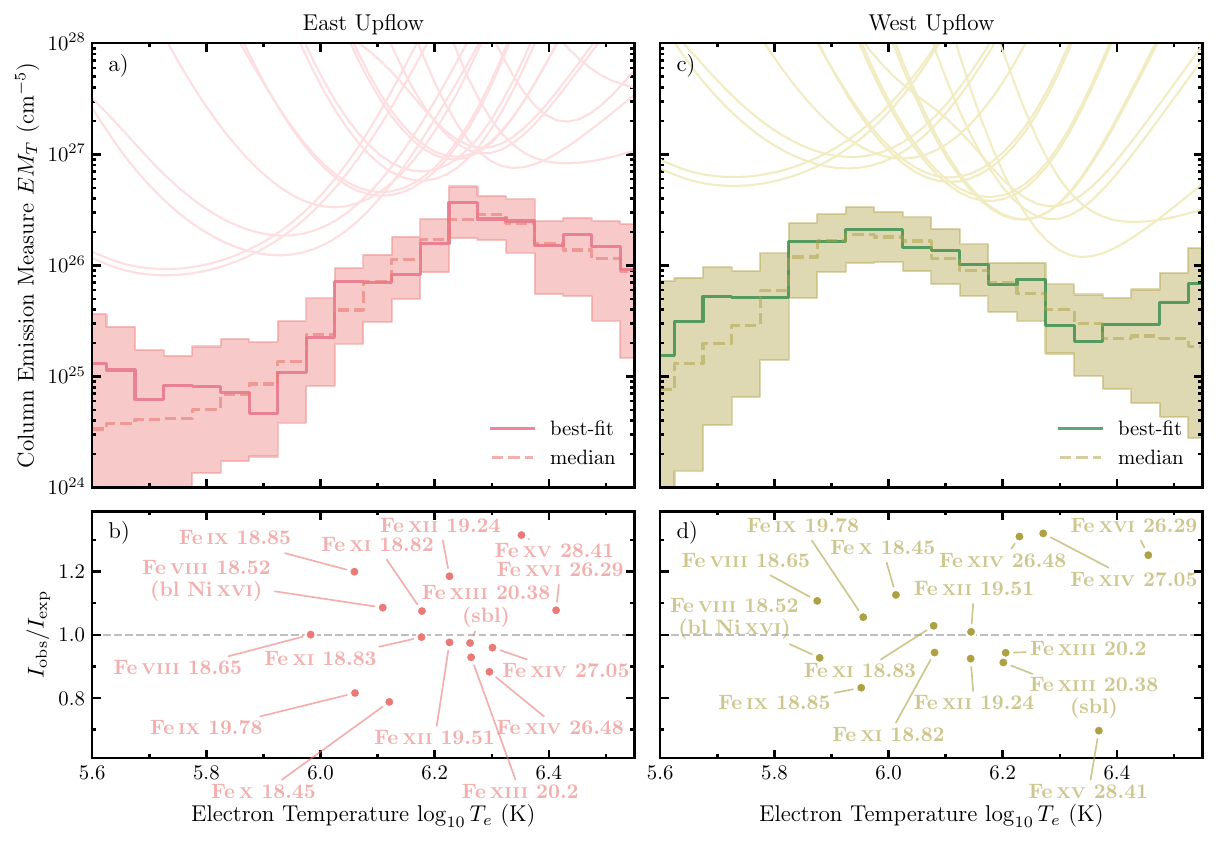}
    \caption{Differential emission measure (DEM) diagnostics of the two upflow regions. Panels (a) and (b) show the results in the eastern upflow region, and Panels (c) and (d) are from the western region (see contours in Figure~\ref{fig:eis_spice_doppler}). Column emission measure $EM_T(T) = DEM(T) \Delta T$ was inferred by an MCMC approach. The upper panels show the best-fit $EM_T$ in solid lines and the median $EM_T$ of 1,000 batches in dashed lines. The shaded areas outline the confidence bounds of the MCMC inference. The EM loci ($I_{\rm obs}/G(T_e)$) curves of the input lines are also shown above the $EM_T$ curves. The bottom panels show the ratios between the observed line intensity $I_{\rm obs}$ and expected line intensity $I_{\rm exp}$ modeled by the best-fit DEM, as a function of the DEM-weighted effective formation temperature. Link to the \texttt{Jupyter} notebook creating this figure: \href{https://yjzhu-solar.github.io/EIS_DKIST_SolO/eis_eui_upflow_ipynb_html/dem_compare.html}{\faBook}.}
    \label{fig:eis_dem}
\end{figure*}

\subsection{Thermodynamic properties}
Observations made by EIS and SPICE reveal different behaviors in the variation in Doppler shifts with temperatures in the eastern and western upflow regions. To explore the thermodynamic structures hosting temperature-dependent Doppler shifts, we performed differential emission measure (DEM) diagnostics on the averaged line intensity (in contours outlined in Figure~\ref{fig:eis_spice_doppler}) using the \texttt{mcmc\_dem} routine in the \texttt{PINTofALE} package \citep{Kashyap1998, Kashyap2000}. The line emissivities were calculated by the CHIANTI atomic database version 10.1 \citep{Dere1997, Young2016, DelZanna2021, Dere2023}. We adopted the default coronal abundance recommended by CHIANTI 10.1, where the abundances of all low-FIP elements increase by 0.5\,dex (a factor of 3.16) compared to the photospheric values recommended by \citet{Asplund2021}. Since we only used low-FIP iron lines in the DEM inversion, the influence of abundance is minimal (due to the blended lines from other elements).  

To calculate the line contribution functions $G(n_e, T)$ required by DEM analysis, we measured the coronal electron number density $n_e$ from intensity ratios between Fe\,\textsc{xiii} 20.20\,nm and Fe\,\textsc{xiii} 20.38 nm (blended) lines. We obtained electron densities of $5.0\times10^8$\,cm$^{-3}$ in the eastern upflow region and $3.1\times10^8$\,cm$^{-3}$ in the western upflow region, which is lower than the typical densities above $10^9$\,cm$^{-3}$ near the footprints of warm AR loops \citep[e.g.,][]{Tripathi2009, Brooks2012, Gupta2015}. However, the values are either close to the density in quiet Sun \citep[$\log N_e \sim 8.5$;][]{DelZanna2012b} or density along an extended fan loop, about 20\arcsec away from its footpoint \citep{Young2012}. The relatively low density in both upflow regions might result from the location of upflows, the decay of the target AR, and the large spatial binning of line profiles across the upflow regions.

The DEM inversion results are summarized in Figure~\ref{fig:eis_dem}, as well as ratios between the observed line intensities $I_{\rm obs}$ and expected line intensities $I_{\rm exp}$ given by the best-fit DEM. Most $I_{\rm obs}/I_{\rm exp}$ ratios are between 0.7 and 1.3, suggesting convincing MCMC DEM inversion results. In the eastern upflow region, the column emission measure $EM_T(T) = DEM(T) \Delta T$ peaks at 1.8\,MK, likely originating from diffuse structures observed in the Fe\,\textsc{xii} intensity map (see Figure~\ref{fig:eis_spice_doppler}). However, the maximum column DEM in the western region appears around 1\,MK, close to the typical formation temperature of fan-like loops. Furthermore, the DEM of the western region shows a broader distribution, revealing an enhanced DEM between 0.6 and 1.0\,MK. 

\begin{figure}[htb!]
    \centering
    \includegraphics[width=\linewidth]{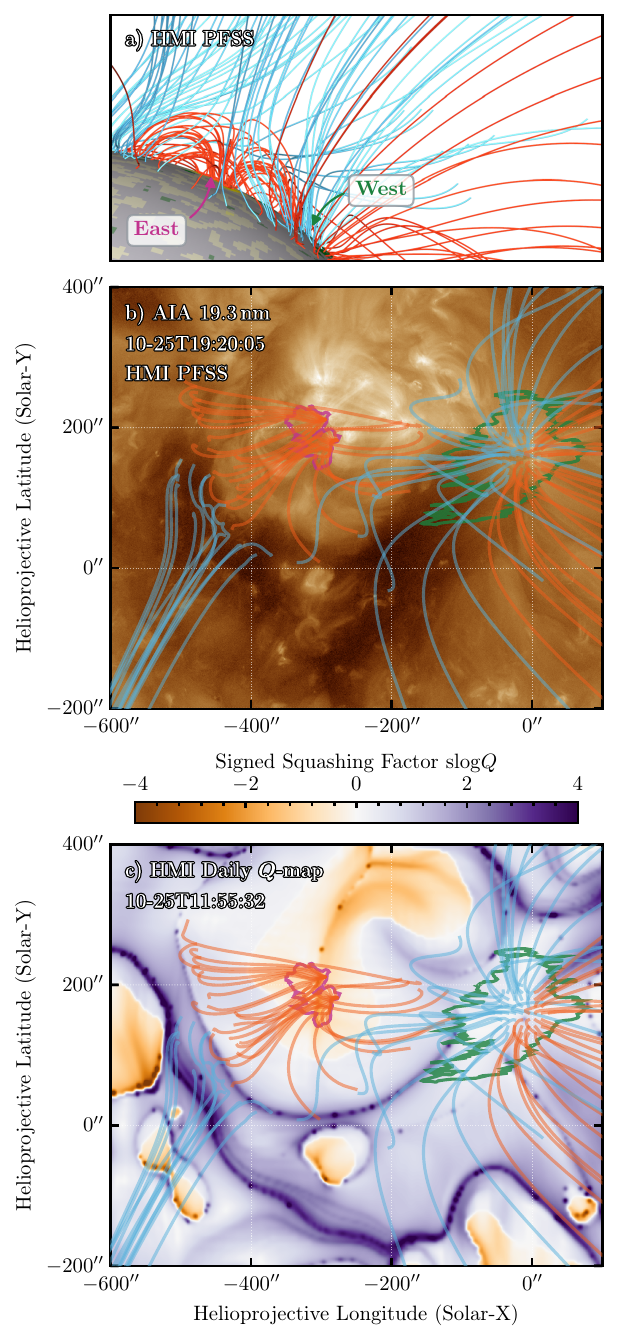}
    \caption{Magnetic field lines traced from the upflow regions using PFSS extrapolation. Open field lines are in blue, while closed field lines are colored red. (a) A side view of the field lines originated in the AR and ambient coronal holes. (b) Extrapolated field lines from the eastern (purple) and western (green) upflow regions and the ambient coronal hole with SDO/AIA 19.3\,nm image. (c) Extrapolated field lines overplotted on the SDO/HMI signed squashing factor $Q$ map. Link to the \texttt{Jupyter} notebook creating this figure: \href{https://yjzhu-solar.github.io/EIS_DKIST_SolO/eis_eui_upflow_ipynb_html/aia_eui_pfss_1025.html}{\faBook}.}
    \label{fig:upflow_pfss}
\end{figure}

\subsection{Magnetic field configuration}

The differences in temperature-dependent Doppler shifts and DEM distributions between the eastern and western regions raise questions about their relationship to different coronal environments, such as distinct magnetic structures. To explore the magnetic field configurations in the upflow regions, we extrapolated the coronal magnetic field using the Potential Field Source Surface \citep[PFSS;][]{Altschuler1969, Schatten1969} extrapolations with polar-filled HMI synoptic magnetograms as the bottom boundary. The side view of the extrapolated field lines from the AR and ambient coronal holes form a pseudostreamer (fan-spine) configuration. The closed field lines in the AR core are surrounded by the open field lines originating from the three low-latitude coronal holes. The eastern upflow region is located between two sets of closed field lines in the center of the pseudostreamer (central spine), whereas the western upflow region lies at the boundary of closed and open field lines (fan-separatrix). 

The traced magnetic field lines from the upflow region are overplotted on AIA 19.3\,nm images, and HMI daily squashing factor $Q$ maps\footnote{\url{http://hmi.stanford.edu/QMap/}} at 1.001\,$R_\odot$ in Figure~\ref{fig:upflow_pfss}. The field lines from the western upflow region form either large transequatorial loops or open field lines connected to the source surface. The open and closed field boundaries are not well resolved in the daily Q-map, probably due to differences in the extrapolated open and closed fields. In contrast, relatively short and closed field lines were found in the eastern upflow region. The other footpoints of these closed field lines are connected to the boundary between the AR and the coronal hole, outlined by the high Q-values.

The magnetic configuration and upflow locations appear to be consistent with the interchange reconnection model suggested by \citet{DelZanna2011}, suggesting pressure-driven upflows caused by reconnection between overpressure AR loops and ambient underpressure field lines.

\subsection{Fine structures in upflow regions}\label{subsec:fine_structures}

The field extrapolation results support the interchange reconnection between close loops in the AR core and ambient open field lines as one of the potential driving mechanisms of both upflow regions. On the other hand, the contributions from small-scale heating in the chromosphere (e.g., type II spicules) or the corona (e.g., nanoflares) have not yet been well discussed in our study. Additionally, due to the limited spatial resolution of spectroscopic observations, Doppler shifts and DEMs are studied over the entire region, lacking insights on small-scale structures in upflows. To address this, we utilized the high-resolution imaging data from HRI\textsubscript{EUV} and IRIS/SJI, revealing a variety of different small-scale dynamics in the transition region, and the lower corona of both upflow regions.  

\subsubsection{Dynamics in the eastern upflow region}

\begin{figure*}[htb!]
    \centering
    \includegraphics[width=\linewidth]{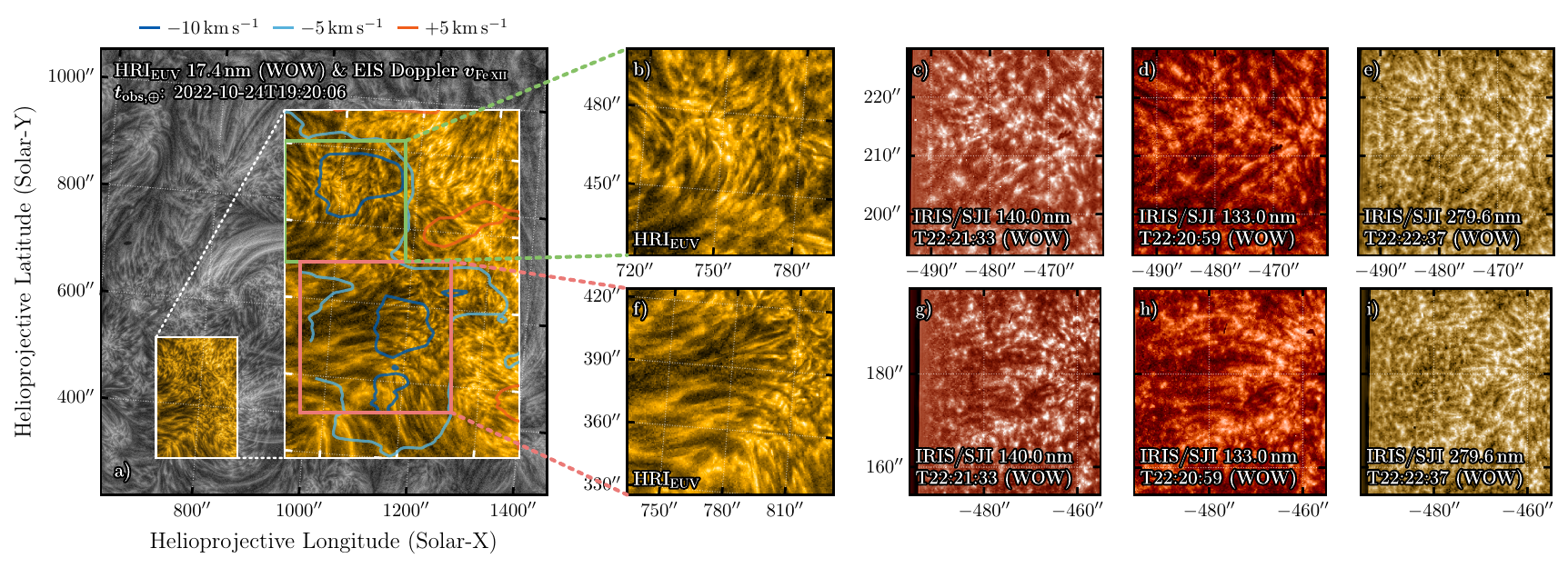}
    \caption{Eastern upflow region observed by HRI\textsubscript{EUV} (WOW-enhanced) and IRIS/SJI on 2022 October 24 (\textbf{Online Movies 1 and 2}). (a) Zoom in plot of the eastern upflow region with EIS Doppler velocity contours overplotted (legends above Panel a). (b) HRI\textsubscript{EUV} image of the top subregion. Panels (c)--(e) IRIS/SJI images of the top subregion in different channels. Panels (f)--(i) are the same as Panels (c)--(e) but for the bottom subregion. The evolution of dynamic fibrils is available as online movies 1 and 2. Link to the \texttt{Jupyter} notebook creating this figure: \href{https://yjzhu-solar.github.io/EIS_DKIST_SolO/eis_eui_upflow_ipynb_html/hri_zoomin_east.html}{\faBook}.}
    \label{fig:hri_east}
\end{figure*}

Figure~\ref{fig:hri_east} shows the clusters of moss- or fibril-like structures with apparent lengths ranging from approximately 1 to 5\,Mm in two subregions of the eastern upflows. Besides, several more elongated fibrils with apparent lengths $>10$\,Mm were observed in the southern subregion. The upflow region appears fainter compared to the AR moss, which is consistent with our DEM analysis. Occasionally, the fibril motion is accompanied by local brightenings with sizes of 0.5 Mm to a few megameters and durations from 1 minute to a few minutes.    

The moss-like structures show both longitudinal and multi-directional motions, similar to the ambient bright AR moss (see Movie 1), which could also be signatures of kink waves in the moss \citep{Morton2014b} and potentially related to the longitudinal and transverse displacement in the chromospheric spicules \citep[e.g.,][]{DePontieu2007a} and AR fibrils \citep[e.g.,][]{Kuridze2012, Morton2014a}.

Although the IRIS/SJI and HRI\textsubscript{EUV} observations were not strictly co-temporal, similar structures in the eastern upflow region can be identified. For example, brightenings in the transition region and chromosphere at the bottom of these fibrils (see Movie 2). In addition, some elongated fibril-like structures were also seen in the IRIS/SJI 133.0\, and 140.0\,nm images, which implies that some fibrils in HRI\textsubscript{EUV} may either correspond to their counterparts at lower transition region temperatures or represent lower-lying features at low transition region temperatures, like some EUV brightenings \citep{Dolliou2024}.

Examples of small-scale dynamics in HRI\textsubscript{EUV} observations of the eastern upflow regions are outlined by stack plots and the normalized standard deviations in the upper half of Figure~\ref{fig:eui_stackplot}. Slits S1--3 are placed in the eastern upflow region observed on 2022 October 20, when a one-hour sequence of HRI\textsubscript{EUV} image was available. Periodic parabolic features can be seen in stack plots, revealing the back-and-forth motions of hot ``dips'' of dynamics fibrils along the slit directions \citep{Mandal2023a}.

\subsubsection{Dynamics in the western upflow region}

\begin{figure*}[htb!]
    \centering
    \includegraphics[width=\linewidth]{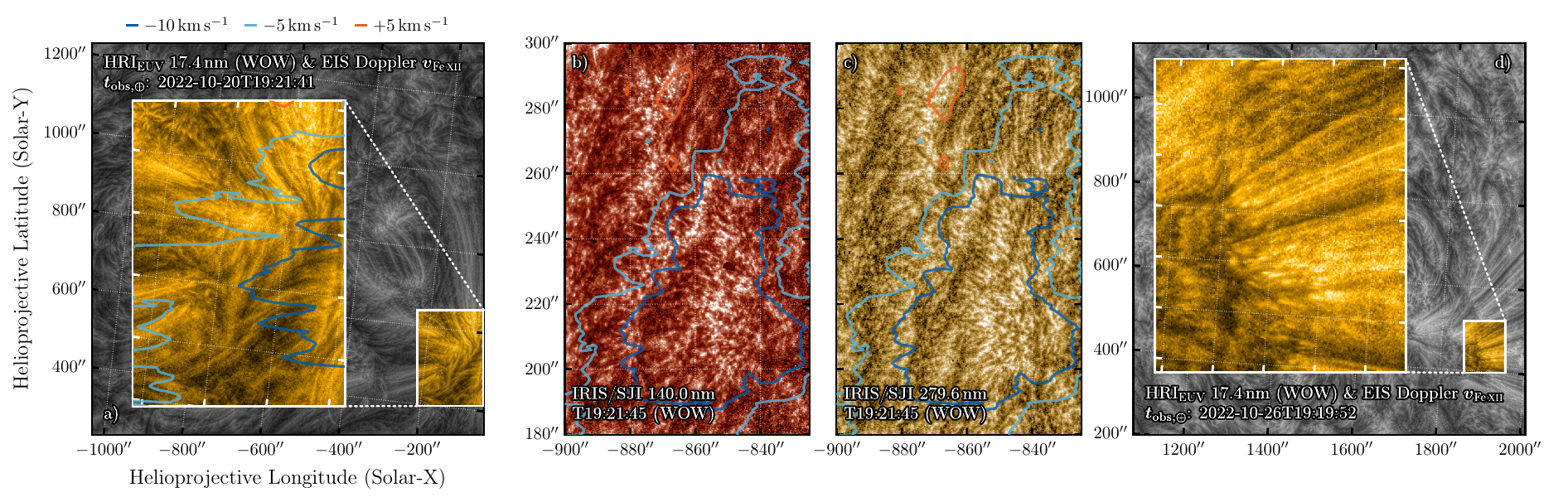}
    \caption{Western upflow region in the fan-like loops (\textbf{Online Movies 3 and 4}). (a) Zoomed-in plot of the upflow region observed by HRI\textsubscript{EUV} on 2022 October 20 with EIS Doppler velocity contours overplotted. (b) and (c) IRIS/SJI images of the western upflow region on 2022 October 20 with EIS Doppler velocity contours. (d) HRI\textsubscript{EUV} image of the western upflow region on 2022 October 26. The evolution of fine structures is available as online movies 3 and 4. Link to the \texttt{Jupyter} notebook creating this figure: \href{https://yjzhu-solar.github.io/EIS_DKIST_SolO/eis_eui_upflow_ipynb_html/hri_zoomin_west.html}{\faBook}.}
    \label{fig:hri_west}
\end{figure*}

Fine structures in the western upflow region are shown in Figure~\ref{fig:hri_west} (and online movies 3 and 4). The western upflow was only well observed by HRI\textsubscript{EUV} and IRIS/SJI when the AR was close to the limb. Unlike the eastern upflow region, the WOW-enhanced HRI\textsubscript{EUV} images reveal multiple strands of fan-like loops. In addition to dynamic fibrils in the background, we found localized EUV brightenings and eruptive jetlet-like features in loop footpoints. The motions of jetlets and dynamic fibrils were generally aligned with the nearby loop strands. 

\begin{figure}[htb]
    \centering
    \includegraphics[width=\linewidth]{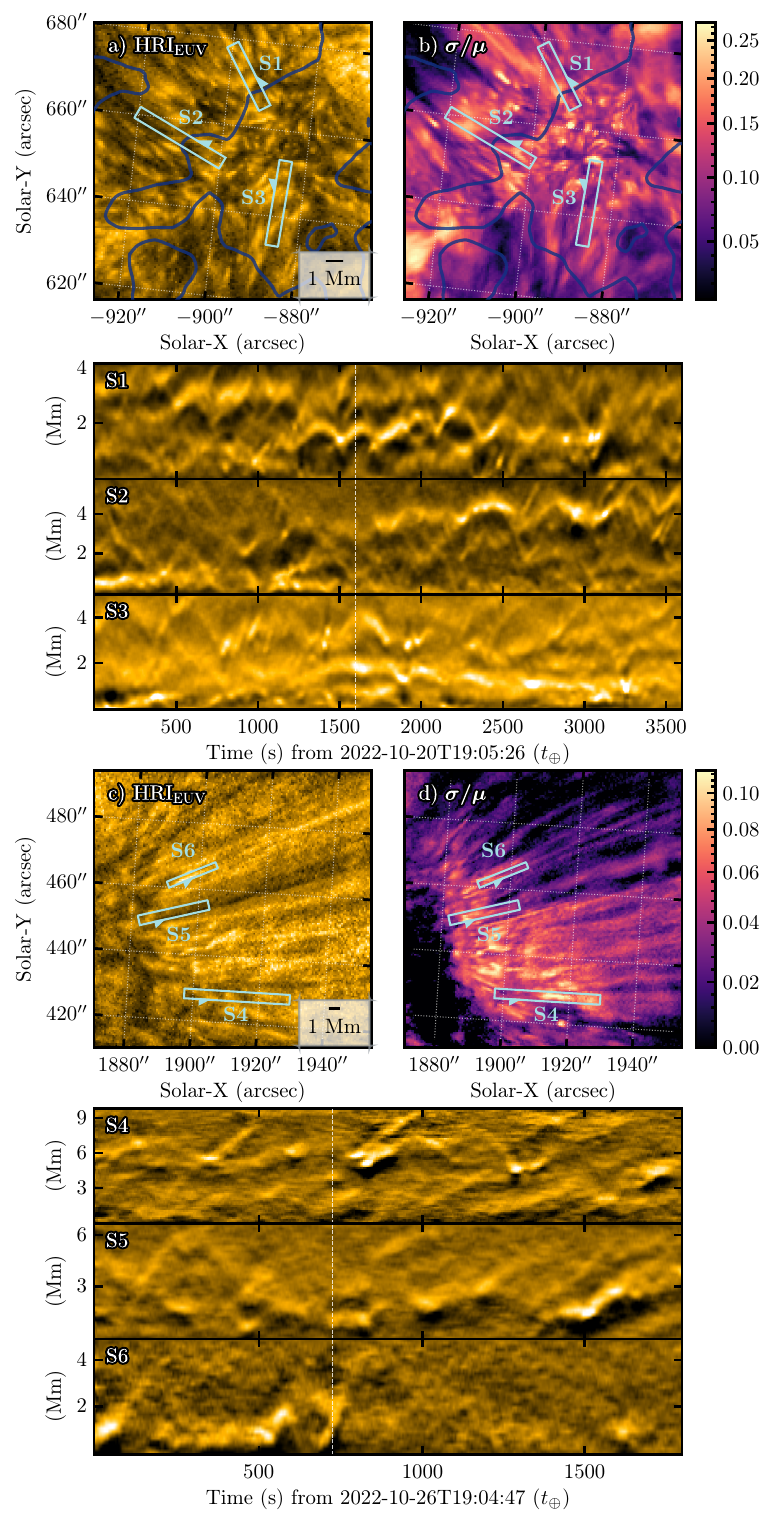}
    \caption{Small-scale dynamics in both upflow regions revealed in standard deviation maps and stack plots. (a) WOW-enhanced HRI\textsubscript{EUV} image and artificial slits S1, S2, and S3 in the eastern upflow region on 2022 October 20. (b) Standard deviations of HRI\textsubscript{EUV} image normalized by average. (S1--3) stack plots along slits S1, S2, and S3. Panels (c) and (d) are the same as Panels (a) and (b), respectively, but for the western upflow region on October 26. Panels (S4--6) are stack plots along the slits S4--6 in Panels (c) and (d). The vertical lines in the stack plots represent the observation time of HRI\textsubscript{EUV} images in panels (a) and (c). Link to the \texttt{Jupyter} notebook creating this figure: \href{https://yjzhu-solar.github.io/EIS_DKIST_SolO/eis_eui_upflow_ipynb_html/eui_stackplot.html}{\faBook}.}
    \label{fig:eui_stackplot}
\end{figure}

The stack plots in the western upflow region on 2022 October 26 are shown in the bottom half of Figure~\ref{fig:eui_stackplot}. Compared to the eastern upflow region, fewer dynamic fibrils were found in the western upflow region S4--6, rather than a chain of periodic dynamic fibrils. This might be caused by projection effects or the decrease in S/N at the detector edge due to vignetting. However, some quasi-periodic motions are still evident in the online movies. Some tiny jetlets were observed in the S4 and S6, with projected lengths of a few megameters and proper motions of approximately 20--40\,km\,s$^{-1}$. The apparent motions appear to be relatively slow, compared to other jetlet-like structures, e.g., Hi-C jetlets \citep[20--110\,km\,s$^{-1}$;][]{Panesar2019} or network jets \citep[80--250\,km\,s$^{-1}$;][]{Tian2014}. Furthermore, these jetlets occupy only a limited fraction of the coronal upflow area at any given time, implying that they may not be major contributors to the entire upflows.

\subsection{Lower atmosphere}
The fine structures revealed by HRI\textsubscript{EUV} and IRIS/SJI in the upflow regions (see Section~\ref{subsec:fine_structures}) suggest a potential connection between the upflow regions and the lower atmosphere. This connection may be essential in determining whether the lower atmosphere actively contributes to driving a significant fraction of the upflowing plasma. To explore the properties of the solar atmosphere below the AR upflow, we analyzed line profiles in the transition region and chromosphere observed by IRIS and CHASE on 2022 October 25. We fitted or inverted them to compare properties such as Doppler shifts and line broadening between the upflow regions and other AR structures.  

\begin{figure*}[htb!]
    \centering
    \includegraphics[width=\linewidth]{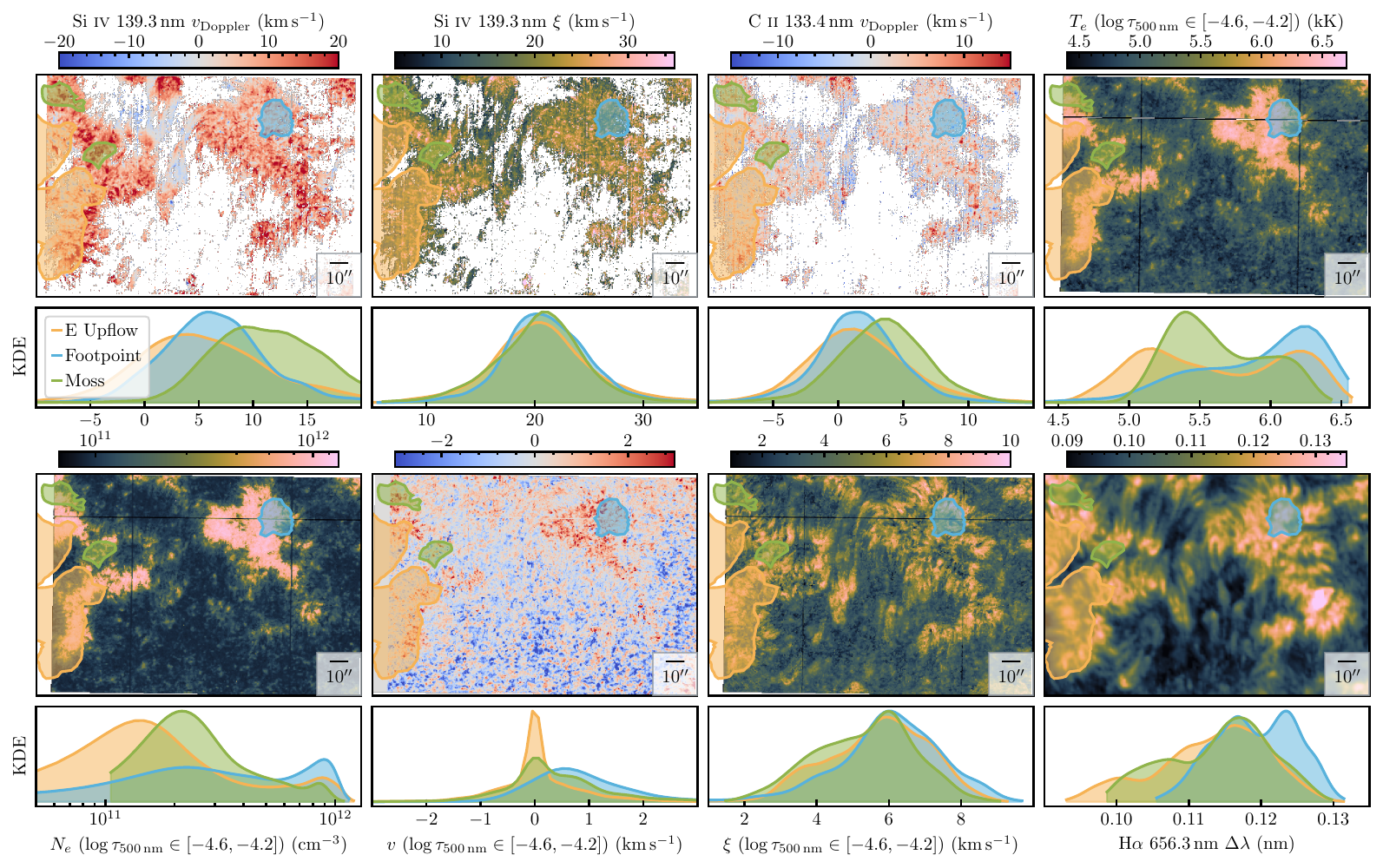}
    \caption{Line Doppler shifts and other thermodynamic properties in the eastern upflow region (orange), footprints of closed loops with blueshifts of approximately $-5$\,km\,s$^{-1}$ in Fe\,\textsc{xii} (blue), and AR moss with redshifts of approximately 5\,km\,s$^{-1}$ in Fe\,\textsc{xii} (green). Top row from left to right: Si\,\textsc{iv} 139.3\,nm Doppler velocity; Si\,\textsc{iv} 139.3\,nm nonthermal velocity $\xi$; C\,\textsc{ii} 133.4\,nm Doppler velocity; IRIS\textsuperscript{2} inverted electron temperature $T_e$ averaged between $\log \tau_{500\,\mathrm{nm}}$ from $-4.6$ to $-4.2$ where Mg\,\textsc{ii} h\&k lines form. Bottom row from left to right: Average inverted electron density $N_e$; Average inverted Doppler shift; Average inverted nonthermal velocity $\xi$; H$\alpha$ line core width $\Delta \lambda$. Link to the \texttt{Jupyter} notebook creating this figure: \href{https://yjzhu-solar.github.io/EIS_DKIST_SolO/eis_eui_upflow_ipynb_html/iris_chase_statistics.html}{\faBook}. }
    \label{fig:iris_stat}
\end{figure*}

Figure~\ref{fig:iris_stat} shows the fitting results and KDEs for three regions of interest: the eastern upflow region, AR moss with redshifts greater than 5\,km\,s$^{-1}$, and a footpoint of a closed loop system characterized by blueshifts of approximately $-5$\,km\,s$^{-1}$. The regions are outlined by contours with the same colors as the corresponding KDE curves. 

In the lower transition region, Si\,\textsc{iv} and C\,\textsc{ii} emission in the three regions is still dominated by redshifts, with a single peak distribution. However, the medians of Si\,\textsc{iv} Doppler shifts gradually decrease from a redshift of around 10\,km\,s$^{-1}$ in the AR moss to less than 5\,km\,s$^{-1}$ in the eastern upflow region. Doppler shifts of C\,\textsc{ii} show a similar pattern among the three regions but with smaller variations from about 5 to 0\,km\,s$^{-1}$. A similar trend in the Doppler distribution was found by \citet{Polito2020}. We note that one of the upflow regions studied by \citet{Polito2020} is also located in a low-intensity region, where fan-like loops were only observed at its edge. The blueshifted tails in Si\,\textsc{iv} and C\,\textsc{ii} Doppler shift distributions suggest a scattered chromospheric population in which both blue shifts are observed in the lower transition region and in the corona \citep[e.g.,][]{Barczynski2021, Huang2021}. Besides, no obvious differences are found in the Si\,\textsc{iv} nonthermal velocity $\xi$ of the three regions. Furthermore, there is no significant correlation among the Si\,\textsc{iv} intensity, Doppler shifts, or nonthermal width in the eastern upflow region, which implies that the blueshifted patches are not necessarily associated with bright transition structures in plages. 

The three regions of interest appear to be located above a dense and hot chromospheric plage region, as indicated by the IRIS\textsuperscript{2} inversion results of the Mg\,\textsc{ii} k line and the H$\alpha$ line widths observed by CHASE. The average temperature and electron density distribution in the upper chromosphere between $\tau_{500\,\mathrm{nm}} = -4.6$ and $-4.2$ show a bimodal distribution, with the hotter and denser component originating from the center of the plage region. The hotter component behaves similarly across the three regions, showing a typical temperature of 6,300\,K, and density of $9\times10^{11}$\,cm$^{-3}$, consistent with previous IRIS observations of plage \citep{Carlsson2015, delaCruzRodriguez2016}. The medians of the average vertical velocity are around zero in the eastern upflow region and AR moss, while at the closed-loop footpoint, the median Doppler shift is less than 1\,km\,s$^{-1}$. The H$\alpha$ line core width observed by CHASE, as a good indicator of temperature in the middle-upper chromosphere \citep{Molnar2019}, also suggests similar heating below the eastern upflow region and the ambient moss. In contrast, the loop footpoint shows slightly greater H$\alpha$ line widths by approximately 0.05\,nm compared to the other two regions.  

\subsection{Persistent upflows}

\begin{figure*}[htb!]
    \centering
    \includegraphics[width=0.95\linewidth]{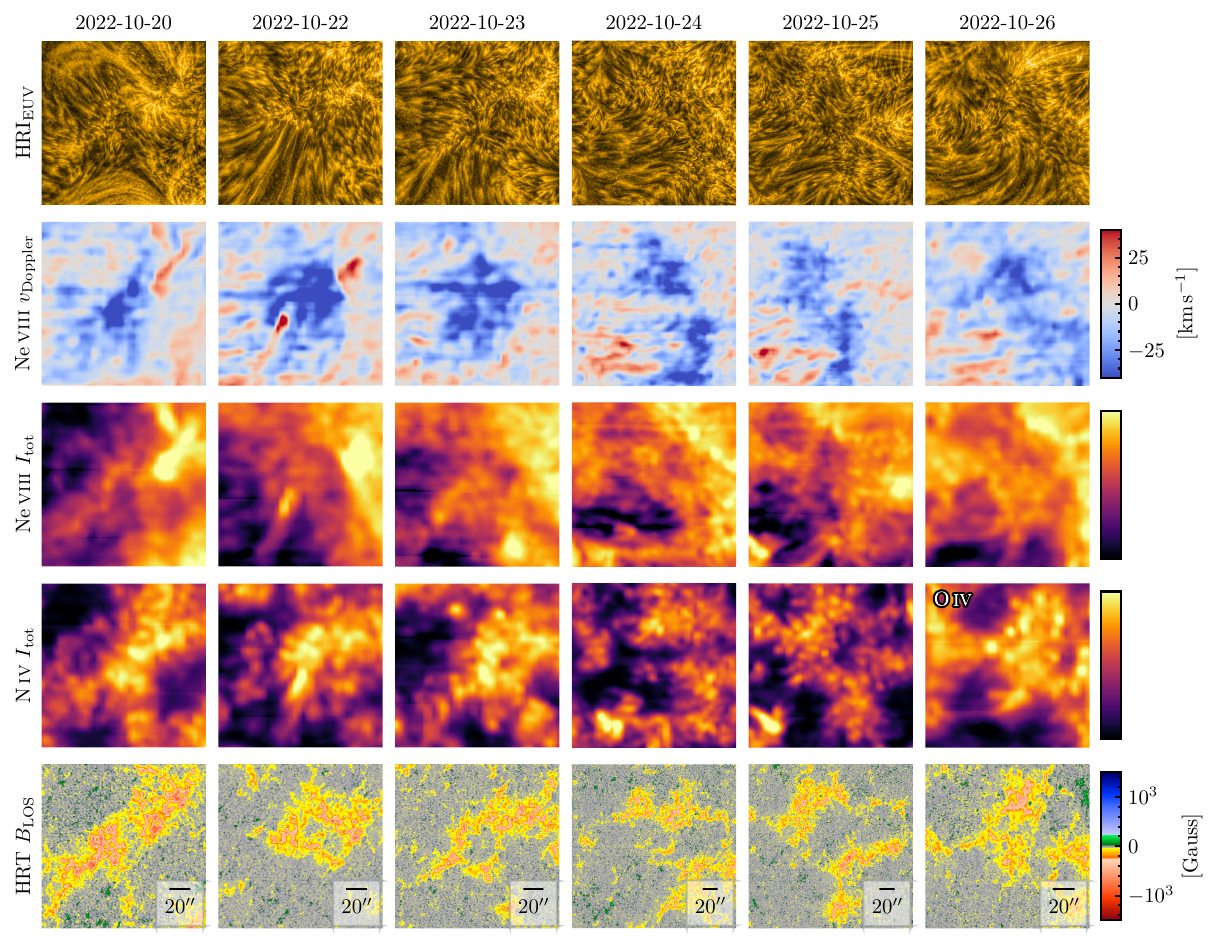}
    \caption{Eastern upflow regions identified by the blueshifts in Ne\,\textsc{viii} 77.04\,nm Dopplergrams observed by SPICE from 2022 October 20 to October 26. First row: HRI\textsubscript{EUV} images; Second row: SPICE Ne\,\textsc{viii} Dopplergrams; Third row: SPICE Ne\,\textsc{viii} intensity maps; Fourth row: SPICE N\,\textsc{iv} 76.51\,nm intensity maps, except for the October 26 one, which was replaced by the O\,\textsc{iv} 78.77\,nm intensity; Last row: PHI/HRT LOS magnetograms. Note that the scale bars vary among different dates. Link to the \texttt{Jupyter} notebook creating this figure: \href{https://yjzhu-solar.github.io/EIS_DKIST_SolO/eis_eui_upflow_ipynb_html/hri_spice_evolution.html}{\faBook}.}
    \label{fig:hri_evo}
\end{figure*}

Previous studies have demonstrated that AR upflows, particularly those associated with fan-like loop structures, can persist for days \citep[e.g.,][]{Demoulin2013}. To investigate whether the properties of the atypical eastern upflow region are persistent or have intermittent behavior, we traced its evolution using instruments on board Solar Orbiter. Figure~\ref{fig:hri_evo} shows the HRI\textsubscript{EUV} intensity, SPICE Ne\,\textsc{viii} intensity and Doppler shifts, and the PHI/HRT LOS magnetograms of the upflows at the eastern boundary between 2022 October 20 and 26. We selected the upflows in Ne\,\textsc{viii} Dopplergrams with spatial sizes of tens of arcseconds to mitigate the influence of spurious Doppler velocity. The upflow region remains present on the eastern edge of the AR, attaching to the bright AR moss from October 20 to 26. However, the morphology and exact location of the upflow region continue to evolve with the unipolar photospheric magnetic flux observed by PHI/HRT.

Moss and fibril-like features, including dynamic fibrils, are consistently observed in the eastern upflow region throughout the campaign in HRI\textsubscript{EUV} images. We did not notice any prominent loops rooted in the eastern upflow region, except on 2022 October 22, when a few strands were seen in the lower left part of the FOV. The corresponding redshifts could be artifacts due to PSF (see examples of spurious velocity in Appendix~\ref{app:spice_psf}). N\,\textsc{iv} intensity maps of the eastern upflow region reveal the plage emission in the lower transition region, while Ne\,\textsc{viii} intensity maps illustrate blurred emission from the AR moss in the upper transition region and lower corona. These results are consistent with previous SUMER observations in AR upflow regions \citep[e.g.,][]{Doschek2006}.

\section{Discussion}\label{sec:dis}

\begin{table*}[htb!]
\caption{Summary of the characteristics of the two AR upflow regions.}  
\label{table:upflow_comp}   
\centering                      
\begin{tabular}{p{0.25\linewidth}p{0.3\linewidth}p{0.3\linewidth}}
\toprule
\toprule
Upflow Properties    & Eastern Upflow Region                     & Western Upflow Region         \\ \midrule
Upflow appears above & 0.6--0.8\,MK (Upper transition region) & $\sim$1\,MK (lower corona) \\
DEM Peaks at         & $\sim$1.5\,MK                          & $\sim$1\,MK                \\
\multirow{2}{*}{Magnetic Field} & \multicolumn{2}{c}{In a pseudostreamer structure}  \\
                     &   Unipolar, closed field lines connected the central spine to coronal hole boundaries     & Unipolar, open field lines or long transequatorial loop near fan-separatrix                \\
Small-scale Dynamics            & Faint moss-like bright dots and fibrils,  no loop structures at 1\,MK & Fan-like loops with a few fibrils and jetlets       \\
Structures Below                & Plage-like, reduced redshifts in Si\,\textsc{iv} compared to mossy plage & Plage-like, with fibrils and spicules                   \\ \bottomrule
\end{tabular}\\
\end{table*}

We conducted a comprehensive study of two upflow regions at the edge of a decaying AR. Notably, the eastern upflow region shows several unique characteristics compared to the western region (as summarized in Table~\ref{table:upflow_comp}). The following discussions will focus on the implications drawn, particularly from the atypical eastern upflow region.

\subsection{Fan-like loops and upflows}
Persistent upflow plasma at AR edges, without corresponding fan-like structures, supports the conjecture that fan-like loops and upflows are associated with two distinct populations of field lines \citep{Warren2011}. An alternative interpretation suggests that these upflows represent the cumulative effect of small-scale dynamics in the AR plage \citep{Brooks2020}. In both cases, AR upflows are suggested to form in various magnetic structures and are not necessarily confined to fan-like loops. While the dynamics and driving mechanisms of fan-like upflow regions have been extensively studied over the past two decades \citep[e.g.,][]{Baker2009, Tian2011b, DelZanna2011, Bryans2016}, investigations of these atypical AR upflow regions offer new insights into the nature of AR upflows. For example, the presence of closed and relatively short field lines challenges the scenario that the upflows are only associated with open magnetic funnels \citep{Marsch2008}. 

\subsection{Temperature dependence of Doppler shifts}

Due to the minimal contamination from fan-like loops, our analysis suggests that net AR upflows may begin to develop in the upper transition region ($T_e \geq 0.6$\,MK), as illustrated by the eastern upflow region. Previous studies \citep[e.g.,][]{Brynildsen1998, Marsch2004,Doschek2006,Scott2013, Polito2020} also observed AR upflows emerging at upper transition region temperatures in faint regions, contrasting with the redshifted fan-like loops. However, earlier studies either focused on ambient fan-like loops or did not compare these blueshifts with the typical redshifts in fan-like upflows below 1\,MK. Moreover, \citet{Demoulin2013} proposed that blueshifted patches in Si\,\textsc{vii} near fan-like loops are actual counterparts of higher-temperature upflows. Our observations of the eastern upflow region, with the absence of fan-like loops, support this scenario, challenging the conclusion that AR boundary upflows primarily originate at coronal temperatures ($\geq 1$\,MK). 

\subsection{Coupled atmospheres}

Our analysis provides compelling evidence that the flows in the lower atmosphere are coupled with and influenced by the upflows in the solar corona. Specifically, this conclusion is supported by: (a) Similar plasma conditions in the chromospheric plage beneath the eastern upflow region and AR moss (see Figure~\ref{fig:iris_stat}). (b) No clear indication of additional heating events in the transition region beneath the eastern upflow region compared to AR moss (increase in bi-directional flows or nonthermal broadening); (c) A gradual reduction in Doppler shift differences between upflows and moss region in the lower atmosphere, consistent with coronal-originating flows;  (d) Unipolar magnetic field, inconsistent with the existence of unresolved lower-lying loops \citep[e.g.,][]{Feldman1983, Dowdy1986} and interchange reconnection-driven flows in the transition region \citep{Tu2005}. Moreover, Si\,\textsc{iv} and C\,\textsc{ii} Doppler shifts beneath the upflows show a single-peak distribution and systematically less redshift compared to the downflows in mossy plage.  Such observations do not support scenarios involving two populations of structures in the transition region, e.g., open funnels and closed loops \citep[e.g.,][]{Tu2005}, or intermittently heated strands \citep{Peter2010}, unless they occur at significantly higher frequencies. In summary, our findings suggest a classical picture of the transition region in the plage beneath the eastern upflow region, where the transition region plasma is connected to the corona both magnetically and thermally \citep[e.g.,][]{Judge2021}.  

Nevertheless, we note that the implication above is limited by this dataset, which lacks (a)  high-cadence photospheric magnetograms, which might miss the small-scale, transient emergence of the opposite polarity in the unipolar plage \citep[e.g.,][]{Chitta2017, Chitta2019}; (b) investigating the potential component reconnection from magnetic braids in plage \citep[e.g.,][]{Bose2024} (c) Doppler shifts measurements between lower ($<0.2$\,MK) and high transition region ($>0.6$\,MK) to reveal the conversion from fragmented patches of blueshifts and redshifts below 0.2\,MK into a net blueshift above 0.6\,MK, which has been a decades-long puzzle. Previous studies revealed that, although the intensity structures show extensive similarities in the lower and upper transition region, the Doppler shift structures shown in the upper transition region are often not well correlated or even show opposite Doppler shifts \citep[e.g.,][]{Marsch2004, Doschek2006}.

\subsection{Roles of small-scale dynamics} \label{subsec:dis_mechanism}

Our analysis does not provide a definitive conclusion on the role of small-scale dynamics in contributing to the upflows, particularly regarding the evidence that supports a coronal origin. A notable challenge is the prevalence of similar fine structures in both the eastern upflow regions and the adjacent mossy plage (also see Online Movies 1--3). If these fine structures universally drive upflows in the lower corona, it remains unclear why similar upflows (tens of kilometers per second) are not consistently observed in the mossy plage. Moreover, if these dynamic fibrils or the associated slow magnetoacoustic waves \citep[e.g.,][]{Verwichte2010} are sources of the blue asymmetries of line profiles, the absence of significant blue asymmetries in the observed moss region becomes puzzling. Potential explanations include: (a) Fine structures do not directly drive the primary component of the AR upflow; (b) Overlying hot and overdense loops above the moss suppress the flow; (c) Emissions from upflows in moss are too weak compared to the bright, stationary plasma above the moss \citep{Doschek2012}; (d) Chromospheric absorption of the moss emission \citep{DePontieu2009b}; (e) The energy flux in underlying dynamics (e.g., fibrils and moss) are much greater than the energy required for driving coronal upflows due to significant density differences. Therefore, it might not be necessary for the upflow to be associated with a specific dynamic feature. 

Furthermore, the IRIS observation revealed no correlation between the intensity and blueshifts in the low transition region (e.g., Si\,\textsc{iv}), which would be expected if chromospheric or low transition region plasma were directly heated and transported into the coronal upflow in these fine structures. Previous studies show conflicting results, e.g., \citet{Huang2021} reported that blueshifted structures are marginally brighter in Si\,\textsc{iv} compared to the redshifted ones, whereas \citet{Feldman2011} found C\,\textsc{iv} redshifts increase with intensities in ARs. Simultaneous high-cadence HRI\textsubscript{EUV} and IRIS dense rasters of upflow regions, avoiding the contamination of fan-like loops, would be of great interest to investigate the relation between the Doppler shifts and fine structures observed in the transition region and corona. Furthermore, a comparison between the Doppler shifts in the bright fibril structures and the dark gaps may become feasible with next-generation EUV spectrographs, e.g., the EUV High-throughput Spectroscopic Telescope (EUVST) on board the Solar-C mission \citep{Shimizu2019}.

\section{Conclusions}
We present a comprehensive study of the upflow region at the edges of a decaying AR, focusing on Doppler shifts and fine structures throughout the solar atmosphere. We identify two upflow regions with Fe\,\textsc{xii} blueshifts exceeding $-10$\,km\,s$^{-1}$. High-resolution HRI\textsubscript{EUV} observations reveal distinct morphologies: the eastern region shows moss-like dynamics, while the western region exhibits fan-like loops. The eastern upflows, uncontaminated by fan-loop emission, provide a unique opportunity to investigate AR upflow characteristics in the corona and underlying atmospheric layers. 

In the western upflow region, blueshifts appear only in spectral lines forming above approximately 1\,MK, whereas lines forming below 1\,MK are dominated by the redshifted emissions from fan-like loops. However, the eastern region shows blueshifts extending to upper transition-region lines forming at approximately 0.6--0.8\,MK (e.g., Fe\,\textsc{viii} and Ne\,\textsc{viii}). We suggest that the eastern region reflects the intrinsic temperature dependence of the Doppler shifts in upflows, i.e., continuous upflows originating from the upper transition region, while the downflows in fan-like loops arise from a different population of field lines unrelated to the upflows. 

Magnetic field extrapolations reveal a pseudostreamer structure above the AR, where the closed field lines from the AR core are enveloped by open field lines originating from the adjacent low-latitude coronal holes. The eastern upflow region is located near the central spine, while the western upflow region is situated near the fan-separatrix. The locations of two upflow regions suggest that flows are driven by pressure imbalances along field lines when dense loops in the AR core reconnect with underdense neighboring loops. 

Observations obtained by HRI\textsubscript{EUV} and IRIS/SJI show numerous moss-like features, fibrils, bright dots, and jet-like fine structures with diverse apparent motions in upflow regions. However, the role of these small-scale dynamics remains inconclusive, as similar small-scale dynamics appear across AR plage without associated upflows. Nevertheless, we confirm the coupling of the lower atmosphere to the coronal upflows and propose that these findings are compatible with a passive response within the magnetically connected transition region to the pressure-driven flows in the corona. 

Our results reveal a complex picture of the AR upflows, where coronal upflows driven by high-altitude reconnection coexist with small-scale dynamics and remain coupled with the underlying transition region. Future investigations should quantify the contributions of small-scale dynamics to the mass flux in upflows and explore how upflow properties relate to their locations within the pseudostreamer configuration (e.g., central spine versus fan-separatrix).

\begin{acknowledgements}
Y.Z. acknowledges support from Karbacher funds and the Swiss National Science Foundation (SNSF) grant number 200021\_219368. Y.Z. also acknowledges the discussion during the ISSI Team 23-585 ``Novel Insights Into Bursts, Bombs, and Brightenings in the Solar Atmosphere from Solar Orbiter,'' led by C. Nelson and L. P. Chitta, especially the helpful suggestions from Z. Huang. Y.Z. would like to thank H. Tian for insightful discussions. L.P.C. gratefully acknowledges funding by the European Union (ERC, ORIGIN, 101039844). Views and opinions expressed are however those of the author(s) only and do not necessarily reflect those of the European Union or the European Research Council. Neither the European Union nor the granting authority can be held responsible for them. P.F.C. was supported by NSFC (12127901).

\newline
Solar Orbiter is a space mission of international collaboration between ESA and NASA, operated by ESA. We are grateful to the ESA SOC and MOC teams for their support. The EUI instrument was built by CSL, IAS, MPS, MSSL/UCL, PMOD/WRC, ROB, LCF/IO with funding from the Belgian Federal Science Policy Office (BELSPO/PRODEX PEA 4000112292 and 4000134088); the Centre National d'Etudes Spatiales (CNES); the UK Space Agency (UKSA); the Bundesministerium f\"ur Wirtschaft und Energie (BMWi) through the Deutsches Zentrum f\"ur Luft- und Raumfahrt (DLR); and the Swiss Space Office (SSO). The development of SPICE has been funded by ESA member states and ESA. It was built and is operated by a multi-national consortium of research institutes supported by their respective funding agencies: STFC RAL (UKSA, hardware lead), IAS (CNES, operations lead), GSFC (NASA), MPS (DLR), PMOD/WRC (Swiss Space Office), SwRI (NASA), UiO (Norwegian Space Agency). The German contribution to SO/PHI is funded by the BMWi through DLR and by MPG central funds. The Spanish contribution is funded by AEI/MCIN/10.13039/501100011033/ and European Union ``NextGenerationEU/PRTR'' (RTI2018-096886-C5,  PID2021-125325OB-C5,  PCI2022-135009-2, PCI2022-135029-2) and ERDF ``A way of making Europe''; ``Center of Excellence Severo Ochoa'' awards to IAA-CSIC (SEV-2017-0709, CEX2021-001131-S); and a Ram\'on y Cajal fellowship awarded to DOS. The French contribution is funded by CNES.
\newline
Hinode is a Japanese mission developed and launched by ISAS/JAXA, collaborating with NAOJ as a domestic partner, NASA and UKSA as international partners. Scientific operation of the Hinode mission is conducted by the Hinode science team organized at ISAS/JAXA. This team mainly consists of scientists from institutes in the partner countries. Support for the post-launch operation is provided by JAXA and NAOJ (Japan), UKSA (U.K.), NASA, ESA, and NSC (Norway). The SDO data are courtesy of NASA/SDO and the AIA, EVE, and HMI science teams. IRIS is a NASA small explorer mission developed and operated by LMSAL with mission operations executed at NASA Ames Research Center and major contributions to downlink communications funded by ESA and the Norwegian Space Centre. This work uses the data from the CHASE mission supported by the China National Space Administration. CHIANTI is a collaborative project involving George Mason University, the University of Michigan (USA), University of Cambridge (UK) and NASA Goddard Space Flight Center (USA). This research has made use of the Astrophysics Data System, funded by NASA under Cooperative Agreement 80NSSC21M00561.
\newline
Packages and software: Astropy \citep{astropy:2013}; Matplotlib \citep{Hunter2007}; Numpy \citep{Harris2020}; Scipy \citep{SciPy2020}; Sunpy \citep{sunpy2020, Sunpy2024_602}; Sunkit-image \citep{Sunkit_image_2023}; Sunkit-magex \citep{Stansby2020}; Sunkit-pyvista \citep{sullivan2019pyvista}; Sunraster; WATROO \citep{Auchere2023}; EISPAC \citep{Weberg2023}; irispy-lmsal; SAFFRON \citep{saffron-spice}; Regions \citep{Regions2023}; Shapely \citep{Shapely2024}; cmcrameri \citep{Crameri2023}; SolarSoft \citep{Freeland2012}; CHIANTI \citep{DelZanna2021, Dere2023}. The \texttt{Jupyter} notebooks and \texttt{IDL} scripts for data reduction, analysis, and visualization are available on Zenodo \citep{ms_code_repo} and GitHub\footnote{\url{https://github.com/yjzhu-solar/EIS_DKIST_SolO}}.

\end{acknowledgements}

\bibliographystyle{aa} 
\bibliography{ms.bib} 

\begin{appendix}

\section{Data calibration and reduction}\label{app:data_calib}
Most imaging data used in this study are either science-ready (e.g., HRI\textsubscript{EUV}) or calibrated using standard routines (e.g., AIA). Spectroscopic datasets processed with additional steps are described in the following subsections. The data from various instruments are coaligned using cross-correlations. Appendix~\ref{app:image_reg} presents details about image alignment and registration. The solar rotation in the long-duration spectroscopic raster-scans has been removed. 

\subsection{EUI}

We analyzed EUI/FSI and EUI/HRI\textsubscript{EUV} L2 data released in EUI data release (DR) 6.0 \citep{euidatarelease6}. FSI data were primarily used in imaging coalignment, while EUI/HRI\textsubscript{EUV} data were used for a detailed analysis of the upflow regions. We removed the image jitter in the HRI\textsubscript{EUV} data by cross-correlation following the procedure described in \citet{Chitta2022} and \citet{Mandal2022}. We made a boxcar average on HRI\textsubscript{EUV} images with a window of 5 frames (30\,s) to increase the S/N. To enhance the visibility of the small-scale UV and EUV structures embedded in these observations, we adopted the wavelet-optimized whitening algorithm \citep[WOW;][]{Auchere2023}. 

\subsection{EIS}
We used the EIS level 1 HDF5 data files prepared by the Naval Research Lab (NRL)\footnote{\url{https://eis.nrl.navy.mil/}} and performed single- or multi-Gaussian fitting using the EIS Python Analysis Code \citep[EISPAC;][]{Weberg2023} package. Each EIS dataset was coaligned with synthetic AIA rasters using the \texttt{eis\_pointing} package \citep[][see more in Appendix~\ref{app:image_reg}]{Pelouze2019}. For absolute wavelength calibration, we adopted the median Doppler shift of each exposure as the velocity zero point. This approach is commonly used in recent EIS datasets \citep[e.g.,][]{Harra2023} to account for the incorrect orbital drift correction after 2018 and the noisy reference Fe\,\textsc{viii} Doppler shifts observed in the quiet Sun \citep{Young2022EISNote16}. We applied the latest onboard radiometric calibration by \citet{DelZanna2025} to the fitted line intensities. Stray light intensities in the upflow regions were removed following the method suggested by \citet{Young2022} (see more discussion in Appendix~\ref{app:stray_light}).

\subsection{SPICE}\label{subsec:spice_method}
The SPICE datasets analyzed in this study are level 2 products from the SPICE data release (DR) 4.0 \citep{SPICEDR4}. The SPICE pointing information in DR 4.0 has a known stellar aberration issue, which causes a shift of tens of arcseconds compared to limb-fitted EUI/FSI images. Therefore, we coaligned SPICE Ne\,\textsc{viii} 77.04\,nm intensity maps with synthetic FSI rasters using routines based on the \texttt{euispice\_coreg} package \citep[][more details in Appendix~\ref{app:image_reg}]{Dolliou2024}. Gaussian fitting of SPICE spectral lines was made by the \texttt{saffron} package\footnote{\url{https://github.com/slimguat/saffron-spice}} with a spatial convolution within a box of 7.7\arcsec\ (7 pixels along the slit) to improve the S/N.

The onboard degradation of the tilted PSF introduces spurious Doppler shifts around bright structures with significant intensity gradients \citep{Fludra2021}. However, we found that consistent Doppler shifts at spatial scales 50--100\arcsec, seen by Solar Orbiter, remain largely unaffected by the PSF artifacts. This is because the sizes of the upflow regions of interest are much larger than the PSF size of 6.7\arcsec\ \citep{Fludra2021}. Furthermore, spatial convolution may also help smooth spurious Doppler signals, as suggested by \citet{Young2012}. Therefore, we present the SPICE Doppler maps generated before the PSF deconvolution in this paper. We emphasize that this study only discusses SPICE Doppler shifts in the upflow regions, where consistent Doppler shifts were observed at spatial scales of tens of arcseconds. We refer interested readers to Appendix~\ref{app:spice_psf} for details on estimations of the spurious velocity and efforts undertaken to apply the PSF deconvolution algorithm \citep{Plowman2023}. 

We initially used the median Ne\,\textsc{viii} 77.04\,nm Doppler shifts in each raster step as the zero point of absolute wavelength calibration. However, most DR 4.0 SPICE raster scans show a quasi-linear drift of Doppler shifts along the raster (west-east) direction. This drift appears to be independent of the heliocentric distances of Solar Orbiter. Although the underlying cause of this drift is still under investigation, we detrended it by fitting a linear function. The final absolute wavelength calibration was made using the median Doppler velocity in the FOV. 

\subsection{IRIS}
We downloaded the IRIS level 2 Slit Jaw Imager (SJI) and raster data from the Heliophysics Events Knowledgebase (HEK). Absolute wavelength calibration of the raster spectrum was made by fitting photospheric lines in IRIS windows. We performed single Gaussian fitting to C\,\textsc{ii} 133.4\,nm (with minimal central reversal in plages) and Si\,\textsc{iv} 139.3\,nm lines and inverted Mg\,\textsc{ii} h\&k lines using the IRIS\textsuperscript{2} code \citep{SainzDalda2019}. 

\subsection{CHASE}
We retrieved CHASE level 1 data from the Solar Science Data Center of Nanjing University. The H$\alpha$ line core (chromospheric) widths were measured by the separation of the half-depth points between the line minimum and line wing intensities at $\pm0.1$\,nm as suggested by \citet{Cauzzi2009} and \citet{Molnar2019}. We further coaligned the limb-fitted CHASE data by comparing the H$\alpha$ core widths with the AIA 160.0\,nm channel, as the H$\alpha$ core widths have a strong correlation with chromospheric heating \citep{Molnar2019}.

\subsection{AIA \& HMI}
Full-disk AIA and HMI L1 data were retrieved from the Joint Science Operations Center (JSOC) and rescaled to a common spatial scale of 0.6\arcsec per pixel. We also utilized the HMI polar-filled synoptic radial magnetograms to perform Potential Field Source Surface (PFSS) extrapolations \citep{Altschuler1969, Schatten1969} and daily Q-maps to investigate the global magnetic field structures associated with the target AR. The daily squashing factor Q maps are part of the HMI synoptic products, based on PFSS extrapolation and a general approach to calculating the Q factor developed by \citet{Titov2008, Titov2011}. 

\subsection{PHI}
We adopted PHI/HRT level 2 LOS magnetograms from the 2\textsuperscript{nd} data release, reduced by HRT pipeline \citep{Sinjan2022} v1.7.0. The Image Stabilization System (ISS) of PHI was turned off during the campaign, and the reconstruction for diffraction at the entrance pupil was not applied in the pipeline. Therefore, the dataset might be influenced by degradation in spatial resolution and polarimetric sensitivity. We coaligned PHI/HRT LOS magnetograms with reprojected HMI magnetograms (see details in Section~\ref{app:image_reg}). 

\section{Image coalignment and registration} \label{app:image_reg}
Coalignments of observations taken by one instrument (e.g., EUI/HRI\textsubscript{EUV}) or several different remote-sensing instruments are important procedures in data reduction, especially for coordinated observation campaigns. Stereoscopic observations from different vantage points make image coalignment and registration more challenging, as one remote-sensing instrument captures 2-D projections of 3-D solar structures. 

In this study, given the typical sizes of tens of arcseconds of the upflow regions determined by a Doppler shift of $-5$\,$\mathrm{km\,s^{-1}}$ in Fe\,\textsc{xii}, we consider a coalignment with an uncertainty of 5\arcsec--10\arcsec\ in Solar Orbiter observations would be sufficient. Besides, EUI/FSI, as a key instrument in coalignments, has a spatial resolution of approximately 10\arcsec. Although sub-pixel optimal shifts can be derived from least-square fitting near the maximum of the cross-correlation coefficients, we estimate that the uncertainty in coalignments between different instruments can be up to around 5\arcsec--10\arcsec, while coalignments within the same instrument may have a lower uncertainty of 2 pixels (1\arcsec\ for HRI\textsubscript{EUV}). 

We coaligned most observations following these main principles:
\begin{enumerate}
    \item Reference coordinates (i.e., update the keywords \texttt{CRVALs} in FITS header) are shifted;
    \item The optimal shift is calculated by cross-correlating two images showing similar features and taken closest in time (e.g., EIS Fe\,\textsc{xii} 19.51\,nm line intensity and AIA 19.3\,nm maps);
    \item Coalignments are usually made in the helioprojective Cartesian coordinates of the maps of which pointings are going to be updated;
    \item If the two images are taken at different times, from different vantage points, or have different pixel scales, we reprojected one image into the other's helioprojective frame using the WCS transformation and Carrington coordinates calculation provided by \texttt{sunpy}, which handles the solar rotation and light travel time correction\footnote{\url{https://docs.sunpy.org/en/stable/topic_guide/coordinates/carrington.html}};
    \item Instruments on Solar Orbiter, if possible, are all coaligned to EUI/FSI, while the near-Earth instruments are coaligned to SDO/AIA or SDO/HMI; Then FSI and AIA are coaligned by cross-correlating their common He\,\textsc{ii} 30.4\,nm channel after reprojection. 
\end{enumerate}

It is worthwhile to mention that the rotation of the image, often represented as the keywords \texttt{CROTAn} or \texttt{PCi\_j} in FTIS headers, although can be poorly recorded, are not updated in most datasets to simplify and boost the coalignment processes. Furthermore, consider an HRI\textsubscript{EUV} image which has a FOV of 1000\arcsec\ (e.g., see Figure~\ref{fig:fov_summary}), a typical tilt of 0.5\degr\ would result in a shift of approximately 9\arcsec\ from the bottom to the top of the FOV, which is still comparable or less than the uncertainty from FSI spatial resolution of 10\arcsec. 

\begin{figure*}[htb!]
    \centering
    \includegraphics[width=0.9\linewidth]{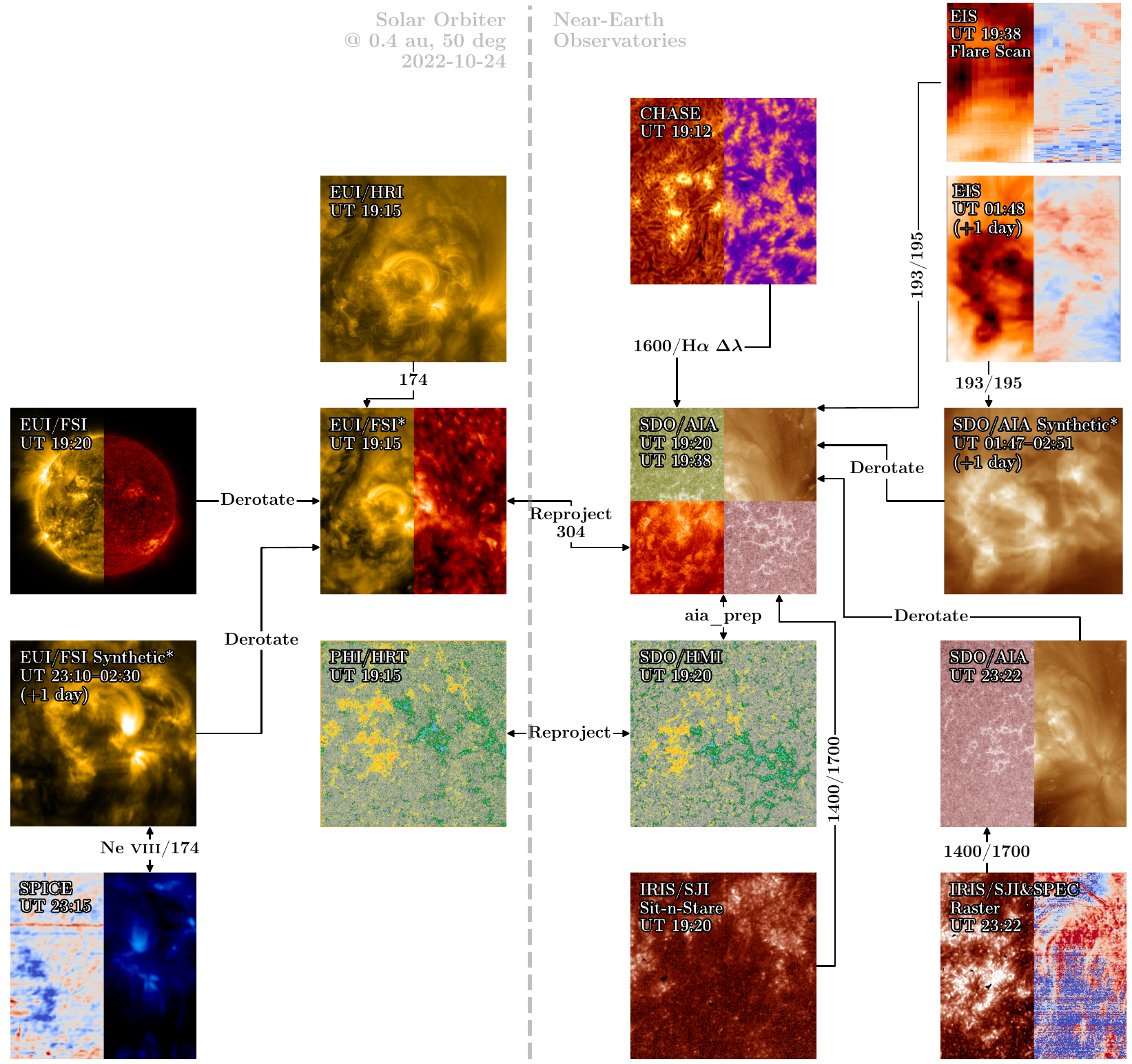}
    \caption{Example of coaligning observations made by Solar Orbiter and other near-Earth observatories on 2022 October 24 and October 25. Arrows connect the maps coaligned with each other. ``Derotate'' means the maps observed at different times are coaligned after the removal of solar rotation. Maps connected by arrows with ``reproject'' undergo projection from one helioprojective frame to another before coalignment. Annotations with slash denote that we cross-correlated the two instruments using these channels or spectral lines; for example, ``1400/1700'' means IRIS and AIA are coaligned by comparing IRIS/SJI 140.0\,nm channel and AIA 170.0\,nm channel. Maps with an asterisk (*) are reprojected maps, either to remove solar rotation or to create a single synthetic raster image from a series of images. Link to the \texttt{Jupyter} notebook creating this figure: \href{https://yjzhu-solar.github.io/EIS_DKIST_SolO/eis_eui_upflow_ipynb_html/coalign_flowmap.html}{\faBook}.}
    \label{fig:app_coalign}
\end{figure*}

Figure~\ref{fig:app_coalign} shows an example of coaligning observations from Solar Orbiter and other near-Earth observatories on 2022 October 24 and October 25. Arrows illustrate how maps are coaligned with each other, ultimately referencing EUI/FSI or SDO/AIA images. Imaging data, e.g., HRI\textsubscript{EUV}, IRIS/SJI, and CHASE, are directly coaligned to FSI or AIA images closest in time. For magnetograms from PHI/HRT, as there is no appropriate reference image available from instruments on board Solar Orbiter (particularly due to the absence of FDT data), we reprojected the HMI map (in a non-flux-conservative manner) to PHI/HRT frames and cross-correlated the reprojected HMI magnetogram with the PHI magnetogram. This method only works for the last few days of the observation campaign when the longitudinal separation between Solar Orbiter and Earth was below around 60\degr). For earlier PHI/HRT observations, we shifted the magnetograms under manual inspection to match the magnetic elements in HRT magnetograms with loop footpoints and local brightenings seen in HRI\textsubscript{EUV}. 

Additionally, we coaligned HRI\textsubscript{EUV} image series following the method described by \citet{Chitta2022}. The HRI\textsubscript{EUV} series was segmented into subgroups of 11 frames, with each subgroup overlapping the subsequent one by a single frame. The frames in each group were cross-correlated internally, while the coalignment between subgroups was established using the overlapped frames. 

Slit spectrograph data obtained by EIS and SPICE were coaligned in special ways, due to the extended period of one or several hours required to finish a large FOV raster. Ideally, one should cross-correlate the 1-D intensity obtained at each exposure with a 2-D image (e.g., SDO/AIA) to determine the accurate pointing of the spectrograph. To minimize the efforts, we adapted the \texttt{eis\_pointing} \citep{Pelouze2019} and \texttt{euispice\_coreg} \citep{Dolliou2024} packages for creating synthetic rasters from AIA and EUI/FSI and cross-correlating EIS and SPICE rasters with the synthetic ones. \texttt{eis\_pointing} builds different synthetic rasters that account for the shift and rotation of EIS maps. On the other hand, we developed our custom routines based on \texttt{euispice\_coreg} to find the optimal rotation matrix (\texttt{PCi\_j}) of SPICE by maximizing the normalized cross-correlation matrix. Additional corrections of SPICE pointing distortion were also applied using \texttt{WCSDVARR} extensions in SPICE L2 files. The outputs from both packages are slightly revised to remove the solar rotation between each exposure.

Additionally, the choice of radial coordinates $r_{\rm repro}$ in reprojection also affects coalignment results, as we inevitably rely on the reprojection of maps from one helioprojective frame to another helioprojective or heliographic frame. Although the actual uncertainty might be less than 0.5\% of the solar radii, legitimate differences in reprojection can be noticed if various $r_{\rm repro}$ are used. For instance, reprojecting an AIA image to HRI\textsubscript{EUV} frames, taken at 0.4\,au with a longitudinal separation of 50\degr, using either $r_{\rm repro} = 1$ or 1.004\,$R_\sun$ (2.8\,Mm above the surface), will result in a shift of approximately 10\arcsec\ in HRI\textsubscript{EUV} frames.  

Unfortunately, there is no single and fixed value of $r_{\rm repro}$ for all channels due to variations in instrumental definitions and formation heights of emissions, especially for extended features in the solar corona, like coronal loops. By definition, EUI adopts the IAU recommended value for the photospheric radius of 695.7\,Mm \citep{Prsa2016} as the value of the keyword \texttt{rsun\_ref}. On the other hand, all SDO/AIA and SDO/HMI images, use a constant reference radius of 696.0\,Mm in their FITS headers. 

Furthermore, the actual formation heights of the transition region and coronal emissions above the photospheric surface could vary in different regions. \citet{Alissandrakis2019} suggested the network emissions in AIA 30.4\,nm channel appear approximately 4\,Mm above the HMI limb in 617.3\,nm and about 3.8\,Mm above the AIA 170.0\,nm emission. The triangulation of EUV brightenings (campfires) in HRI\textsubscript{EUV} channel suggested an average formation height of approximately 2.8\,Mm above the IAU reference $R_\odot$ \citep{Zhukov2021}. Using the time delays between 3-minute oscillations in different AIA channels, \citet{Deres2015} estimated the AIA 17.1\,nm and 19.3\,nm emissions are about 1--3\,Mm above the AIA 170.0\,nm emission in sunspot umbra, while AIA 30.4\,nm forms about 500\,km above the AIA 170.0\,nm. A recent study by \citet{Sanjay2024} suggested AIA 30.4\,nm passband forms around 850\,km above the HMI continuum, while the 17.1\,nm channel forms at around 1500\,km. 

Given all these uncertainties, we adopted a crude simplification to this problem by dividing the reprojected maps into two categories: maps ``below the corona'' and maps ``in the corona.'' Maps classified as ``below the corona'' (e.g., HMI magnetograms, AIA 30.4\,nm, and FSI 30.4\,nm images) are reprojected at $r_{\rm repro} = 696\,$Mm. On the other hand, maps ``in the corona'' like AIA 17.1\,nm and EIS Fe\,\textsc{xii} 19.51\,nm maps are reprojected 2.8\,Mm above the ``surface'' at 696\,Mm. Manual inspection confirmed that this approach provides better coalignment results than unifying all $r_{\rm repro}$ to 696\,Mm. 

\section{SPICE PSF and spurious Doppler velocity} \label{app:spice_psf}
An unexpected onboard PSF degradation of SPICE was noticed early in the mission. \citet{Fludra2021} estimated degradations from 5.4\arcsec\ to 6.7\arcsec\ in the spatial dimension, 4.7 pixels to 7.8 pixels, and 5.3 pixels to 9.4 pixels in the spectral dimensions of SW and LW detectors, respectively. The degraded PSF not only reduces the spatial resolution of SPICE but also influences the measurement of Doppler shifts in spectral lines. 

The tilt of the elliptical SPICE PSF on the slit-aligned and spectral ($y$-$\lambda$) plane introduces spurious Doppler shift signals associated with intensity gradients \citep{Plowman2023}. Similar effects were also found in previous solar UV/EUV spectrographs \citep[e.g.,][]{Haugan1999, Young2012, Warren2018}. However, due to the aforementioned PSF degradation, these spurious Doppler shifts might reach magnitudes of tens of kilometers per second in ARs observed by SPICE \citep[e.g.,][]{Janvier2023, Petrova2024}, which may significantly impact the measurement of Doppler shifts related to AR upflows. 

In addition to these spurious Doppler shifts, other artifacts in SPICE data suggest a 3-D PSF that blends the spectral information across the spatial ($x$, $y$), and spectral ($\lambda$) axes. For example, systematic shifts along both $x$ and $y$ directions are found when comparing the monochromatic intensity maps at different wavelengths. 

\begin{figure}[htb!]
    \centering
    \includegraphics[width=\linewidth]{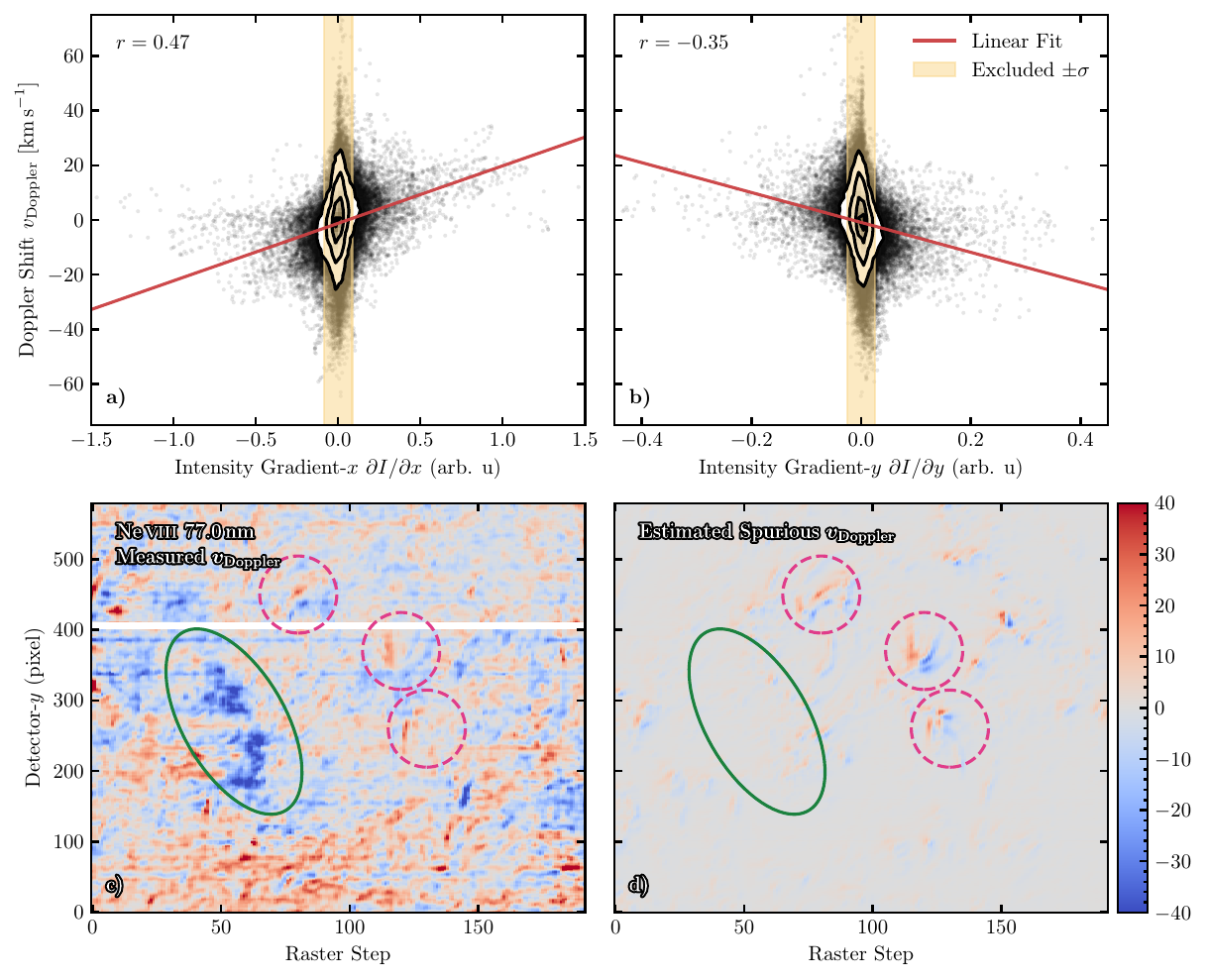}
    \caption{Proxy of spurious velocity in the SPICE Ne\,\textsc{viii} 77.04\,nm line estimated by intensity gradients. Panels (a) and (b) show 2-D histograms of Doppler velocity vs. Intensity gradient along the raster ($x$) and slit ($y$) directions, respectively. Red lines show linear regression fits, with Pearson correlation coefficients $r$ indicated in the upper-left corners. Shaded yellow areas denote pixels excluded from linear regression, having intensity gradients within $\pm 1\sigma$. Panels (c) and (d) compare measured Doppler velocities and proxies of spurious velocity estimated by the regression slopes shown in Panels (a) and (b). The eastern upflow region is highlighted by the green ellipses, whereas dashed magenta circles outline the prominent spurious velocity patterns. Link to the \texttt{Jupyter} notebook creating this figure: \href{https://yjzhu-solar.github.io/EIS_DKIST_SolO/eis_eui_upflow_ipynb_html/spice_spurious_vel.html}{\faBook}.}
    \label{fig:spice_spurious}
\end{figure}

We adopted three methods to evaluate the potential influence of the tilted PSF on Doppler shifts measured in the eastern upflow region. We first estimated the magnitude of spurious velocity using linear regression between Doppler velocities and intensity gradients \citep[also see][]{Janvier2023, Petrova2024}. Pixels with intensity gradients within the $\pm 1\sigma$ interval are excluded from the regression to reduce contamination by actual Doppler signals. A proxy for spurious velocity is given by summing the products of the slopes of two linear regressions $k_x$ and $k_y$ and local intensity gradients, i.e.,
\begin{equation}
    v_{s}(x,y) = k_x \left.\frac{\partial I}{\partial x}\right|_{x,y} +  k_y \left.\frac{\partial I}{\partial y}\right|_{x,y} 
\end{equation}

Figure~\ref{fig:spice_spurious} shows the estimated spurious velocity. The Ne\,\textsc{viii} Doppler shifts exhibit weak or moderate correlations with both the intensity gradient in both the raster ($x$) and slit ($y$) directions, which also implies a more complicated 3-D ($x$-$y$-$\lambda$) PSF. Similar spurious patterns appear in the measured Doppler shifts and the estimated proxies near bright structures (dashed circles). However, we do not find any significant spurious Doppler shift signals in the eastern upflow region, where consistent blueshifts are observed.

As a second approach, we applied the SHARPESST PSF deconvolution scheme developed by \citet{Plowman2023} on 2022 October 24 SPICE datasets, with spatial binning by a factor of 2 along the $y$ axis. A major challenge here is that the instrumental PSF might vary with slits, the short/long-wavelength detectors, and the heliocentric distance of Solar Orbiter, necessitating extensive adjustment in PSF parameters. The eight parameters defining the elliptical $y$-$\lambda$ PSF of Ne\,\textsc{viii} 77.04\,nm line are listed in Table~\ref{table:psf_param}. We first adjust the parameters along the spectral dimension to obtain reasonable profiles at the brightest pixels. Subsequently, we varied the parameters along the spatial dimension and the PSF rotation angle to mitigate spurious velocity patterns in the Dopplergram. Compared to \citet{Plowman2023}, we increase the full widths at half maximum (FWHMs) along the spectral dimension to avoid unrealistic secondary emission peaks at wings. Additionally, we increase the rotation angle to 25\degr\ for better removal of spurious velocities. We acknowledge that the parameters might not be optimal because of (a) the absence of other reference Dopplergrams, (b) the uncertainty in manual parameter tuning, and (c) the degeneracy among PSF parameters.  

\begin{table}[htb!]
\caption{Parameters used in this study defining the Ne\,\textsc{viii} 77.04\,nm line PSF, in comparison to parameters reported by \citet{Plowman2023} for the C\,\textsc{iii} 97.7\,nm line. Note that the parameters used in this study may not be fully optimized. }                 
\label{table:psf_param}    
\centering                        
\begin{tabular}{c c c}      
\hline\hline               
Parameter & This Work & \citet{Plowman2023} \\         
\hline                      
   $F_{y_0,c}$\tablefootmark{(a)} & 2.5\arcsec & 2.0\arcsec \\    
   $F_{\lambda_0.c}$\tablefootmark{(b)} & 1.11\,\AA & 0.95\,\AA    \\
   $\theta_c$\tablefootmark{(c)} & 25\degr & 15\degr or 20\degr \\
   $\gamma_c$\tablefootmark{(d)} & 1.8 & 1.5 \\
   $F_{y_0,w}$\tablefootmark{(e)} & 13.0\arcsec & 10.0\arcsec \\
   $F_{\lambda_0,w}$\tablefootmark{(f)} & 4.5\,\AA & 2.5\,\AA \\
   $\gamma_w$\tablefootmark{(g)} & 1.0 & 1.0 \\
   $w_w$\tablefootmark{(h)} & 0.28 & 0.2 \\
\hline                                 
\end{tabular} \\
\tablefoot{
\tablefoottext{a}{PSF core $y$ axis FWHM before rotation or exponentiation.}
\tablefoottext{b}{PSF core $\lambda$ axis FWHM before rotation or exponentiation.}
\tablefoottext{c}{PSF core rotation angle.}
\tablefoottext{d}{PSF core non-Gaussian exponentiation.}
\tablefoottext{e}{PSF wing $y$ axis FWHM before rotation or exponentiation.}
\tablefoottext{f}{PSF wing $\lambda$ axis FWHM before rotation or exponentiation.}
\tablefoottext{g}{PSF wing non-Gaussian exponent.}
\tablefoottext{h}{Wing weight.}
}
\end{table}

\begin{figure}[htb!]
    \centering
    \includegraphics[width=\linewidth]{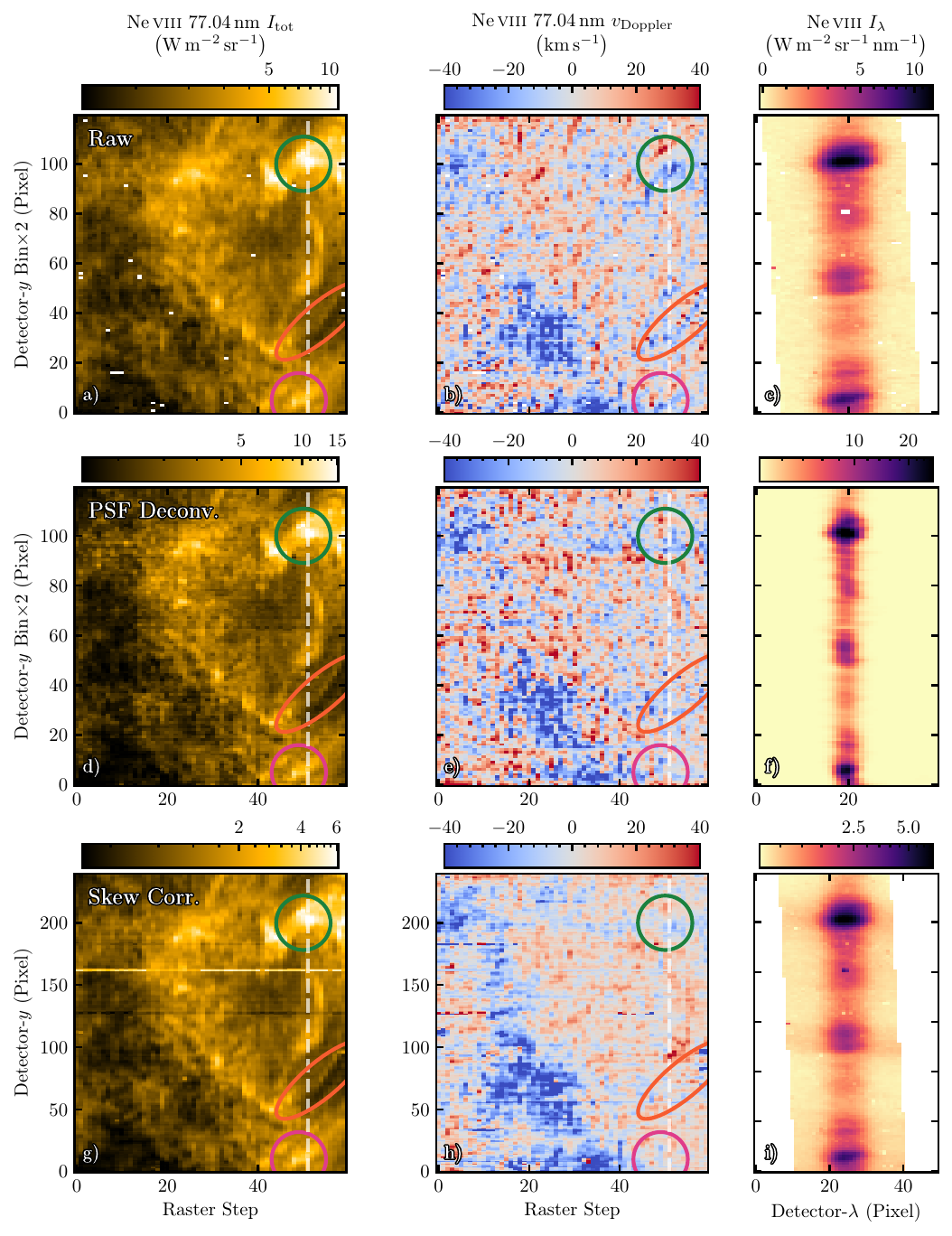}
    \caption{Comparison between the Ne\,\textsc{viii} 77.04\,nm profiles before (top) and after PSF deconvolution (middle) or ``skew'' correction (bottom) near the eastern upflow region. Panels (a), (d), and (g) show Ne\,\textsc{viii} line intensity. Panels (b), (e), and (h) show Doppler shifts. Panels (c), (f), and (i) display spectra along the dashed vertical lines in other panels, respectively. Regions potentially affected by spurious velocity signals are highlighted by circles. Link to the \texttt{Jupyter} notebook creating this figure: \href{https://yjzhu-solar.github.io/EIS_DKIST_SolO/eis_eui_upflow_ipynb_html/spice_1024_NeVIII_ms_version.html}{\faBook}.} 
    \label{fig:spice_psf_deconv}
\end{figure}

The third method for mitigating the PSF artifacts is based on the additional spatial shifts in monochromatic images. Instead of employing 
 computationally expensive 3-D $x$-$y$-$\lambda$ PSF deconvolution, this approach applies a wavelength-dependent shift to monochromatic images\footnote{\url{https://github.com/jeplowman/spice-line-fits}}. The shifts approximate the effect of the tilted 3-D PSF, as a ``skew'' operation in both $x$-$\lambda$ and $y$-$\lambda$ planes. Optimal shift values of 1.667\arcsec\,\AA$^{-1}$ and $-$1.667\arcsec\,\AA$^{-1}$ along the $x$ and $y$ directions were determined by minimizing the standard deviation across the entire Dopplergram using an adaptively refined mesh. Additionally, the distortion caused by the skew correction was removed after the fitting. 

Figure~\ref{fig:spice_psf_deconv} compares the Ne\,\textsc{viii} 77.04\,nm intensity, Doppler shifts, and line profiles before and after applying either PSF deconvolution or the skew correction. Several spurious velocity patterns (outlined by circles) are significantly reduced following correction. The tilted features in Figure~\ref{fig:spice_psf_deconv}c are diminished in Figure~\ref{fig:spice_psf_deconv}f and i. In addition, the skew correction also significantly reduces extrema in the Dopplergrams. Importantly, blueshifted patterns in the eastern upflow region remain largely unchanged by either correction method, confirming that the eastern upflow region is not an artifact caused by the PSF.  

In summary, we conclude that the tilted $y$-$\lambda$ PSF of SPICE does not significantly affect the Ne\,\textsc{viii} blueshifts with spatial scales of tens of arcseconds found in the eastern upflow region. Since the PSF deconvolution and skew correction approaches are experimental, in the main text, we still present the Doppler shifts inferred without applying these corrections.

\section{EIS stray light estimation} \label{app:stray_light}

For the two upflow regions studied in this work, the eastern upflows are very close to the AR moss, while the western upflows in fan-like loops are much fainter in emissions originating in plasma hotter than 2\,MK (e.g., Fe\,\textsc{xiv} and \textsc{xv}). This raises concerns about the potential contamination of line profiles in upflow regions by instrumental stray light, also known as scattered light, which is crucial for Doppler shift measurements and plasma diagnostics relying on EIS observations. 

As an EUV spectrometer designed to minimize the instrumental stray light \citep{Culhane2007}, it was previously estimated that EIS has an off-limb stray light level of 2\% from a partial solar eclipse seen by Hinode\footnote{\url{https://sohoftp.nascom.nasa.gov/solarsoft/hinode/eis/doc/eis_notes/12_STRAY_LIGHT/eis_swnote_12.pdf}}. However, \citet{Wendeln2018} reported that the stray light in an equatorial coronal hole could reach approximately 10\% of the intensity in the surrounding quiet Sun regions. Due to the limited FOV of EIS rasters, \citet{Wendeln2018} did not investigate the stray light contributions from the entire solar disk. 

\begin{figure}[htb!]
    \centering
    \includegraphics[width=\linewidth]{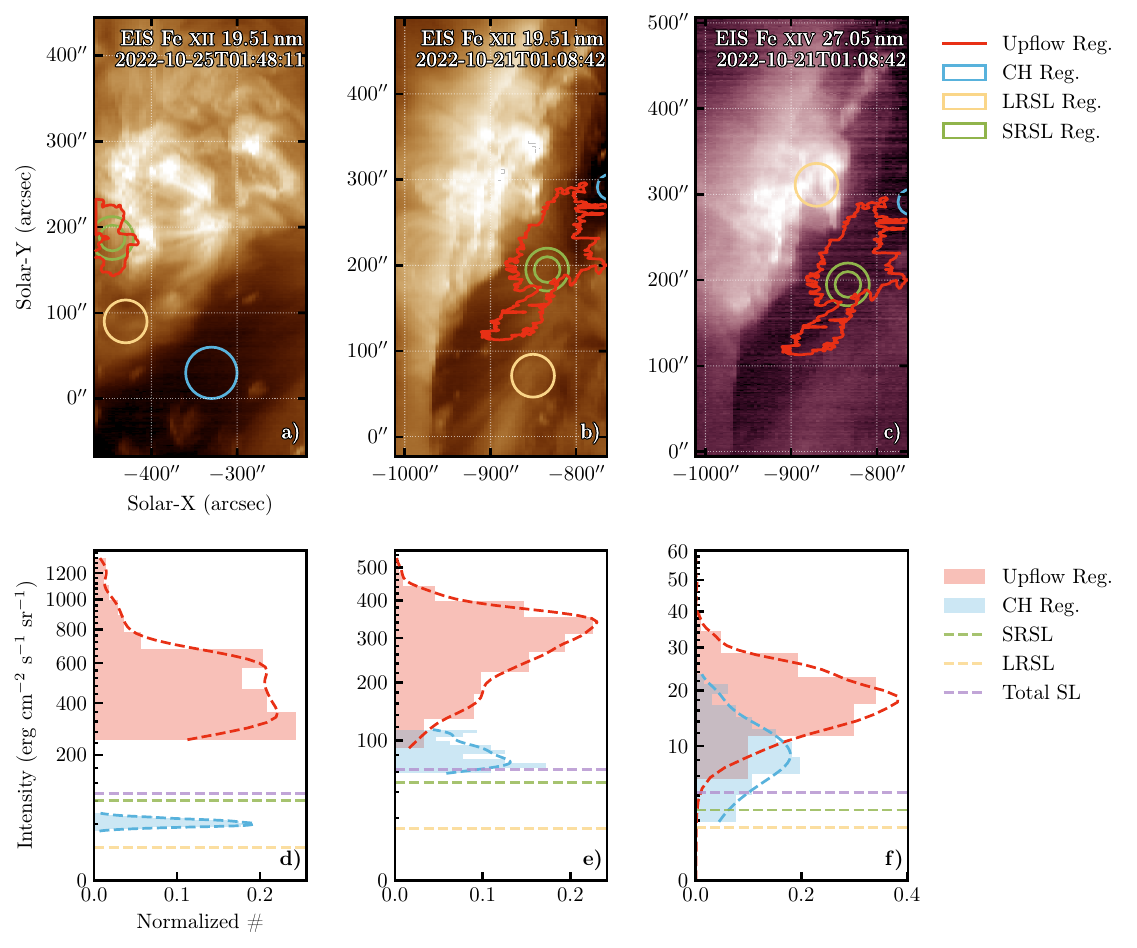}
    \caption{Estimation of EIS stray light intensity in upflow regions. Panels (a)--(c): EIS intensity maps of Fe\,\textsc{xii} 19.51\,nm and Fe\,\textsc{xiv} 27.05\,nm lines. Different colors outline the selected regions: upflows--red polygons; coronal holes--blue circles; long-range stray light (LRSL) reference regions--yellow circles; short-range stray light (SRSL) regions--green annulus. The intensity distributions in upflow regions and ambient coronal holes and estimated short-range stray light and long-range stray light intensities are shown in Panels (d)--(f) below, corresponding to Panels (a)--(c), respectively. Link to the \texttt{Jupyter} notebook creating this figure: \href{https://yjzhu-solar.github.io/EIS_DKIST_SolO/eis_eui_upflow_ipynb_html/app_eis_stray_light.html}{\faBook}.}
    \label{fig:app_eis_stray_light}
\end{figure}

In this study, we adopted a rigorous method recently developed by \citet{Young2022} to separately estimate the short-range stray light (SRSL) and long-range stray light (LRSL) for EIS on-disk observations, with an uncertainty of approximately 25\%. Figure~\ref{fig:app_eis_stray_light} shows examples of stray light estimation in the two upflow regions. The SRSL is estimated by averaging the intensity inside an annulus (green circles in Fig.~\ref{fig:app_eis_stray_light}a--c) centered at the centroid of the upflow region, of which the inner and outer radii are 30\arcsec and 50\arcsec, respectively. On the other hand, the LRSL is estimated as a fraction of the full-disk intensity of the selected EIS line. In practice, an AIA image from a channel with a similar formation temperature (e.g., AIA 19.3 nm--Fe\,\textsc{xii}, and AIA 21.1\,nm--Fe\,\textsc{xiv}) is used as a proxy of the unavailable EIS full-disk intensity map. A reference region (yellow circles in Fig.~\ref{fig:app_eis_stray_light}a-c) is manually chosen to calculate the ratio to convert the AIA map to an EIS full-disk map proxy. The total stray light intensity is removed from the line intensities used for density and DEM diagnostics of both upflow regions. 

For the eastern upflow region (Fig.~\ref{fig:app_eis_stray_light}a and d), the total estimated stray light of Fe\,\textsc{xii} 19.51\,nm is approximately 10--20\% of the values in upflows, contributed mainly by SRSL from the ambient AR core. The estimated stray light intensity is about twice as high as coronal hole intensities, probably because the selected coronal hole region is further away from the AR core compared to the eastern upflow region. The stray light estimation shows more uncertainty in the western upflow region, where the spatial extent of upflows is much greater than the SRSL annulus and does not overlap the bright AR core. For the Fe\,\textsc{xii} 19.51\,nm line, LRSL contributes to about 5\% of the intensity in the upflow region, while SRSL may contribute around 10--15\%. However, in the coronal hole, up to 60--80\% of Fe\,\textsc{xii} 19.51\,nm emission might be stray light. For the Fe\,\textsc{xiv} 27.05\,nm line forming at 2\,MK, coronal hole intensities can be a better estimation of the stray light \citep{Wendeln2018}. Therefore, the stray light may contribute to roughly 50\% of Fe\,\textsc{xiv} emission in the western upflow region. 

We further investigated whether the stray light would affect the Doppler shift measurements in the upflow regions. For the eastern upflow region, the difference between the Doppler shifts of line profiles with or without stray light removal is less than 1.5\,$\mathrm{km\,s^{-1}}$, which is much less than the EIS wavelength calibration uncertainty up to approximately 5--10\,$\mathrm{km\,s^{-1}}$ \citep{Kamio2010, Young2022EISNote16}. Therefore, we concluded that stray light removal is not crucial for accurate Doppler shift measurements with EIS in the upflow regions we investigated. Consequently, we did not remove the stray light when generating Dopplergrams. 

\end{appendix}
\end{CJK*}

\end{document}